\documentclass[fleqn,usenatbib]{mnras}

\usepackage{newtxtext,newtxmath}

\usepackage[T1]{fontenc}
\usepackage{ae,aecompl}

\usepackage{graphicx}	
\usepackage{amsmath}	
\usepackage{multicol}
\usepackage[flushleft]{threeparttable}
\usepackage{supertabular,booktabs}
\usepackage{pdflscape} 
\usepackage{multicol}
\usepackage[usenames,dvipsnames,table]{xcolor}        
\usepackage{colortbl}

\newcommand{\erg}{{\hbox{erg~s$^{-1}$}}}             

\newcommand{\etal}{{{et~al.}}}

\newcommand{\msun}{{\hbox{M$_{\odot}$}}}

\newcommand{\msunpyr}{{\hbox{M$_{\odot}\,\rm{yr^{-1}}$}}}

\newcommand{\Ha}{H$\alpha$}
\newcommand{\Hb}{H$\beta$}

\usepackage{tikz}
\definecolor{lime}{HTML}{A6CE39}
\DeclareRobustCommand{\orcidicon}{%
	\begin{tikzpicture}
	\draw[lime, fill=lime] (0,0) 
	circle [radius=0.16] 
	node[white] {{\fontfamily{qag}\selectfont \tiny ID}};
	\draw[white, fill=white] (-0.0625,0.095) 
	circle [radius=0.007];
	\end{tikzpicture}
	\hspace{-2mm}
}
\foreach \x in {A, ..., Z}{%
	\expandafter\xdef\csname orcid\x\endcsname{\noexpand\href{https://orcid.org/\csname orcidauthor\x\endcsname}{\noexpand\orcidicon}}
}

\title[Comparison of H$\alpha$ and other SFR indicators]{The Star Formation Reference Survey-V: the effect of extinction, stellar mass, metallicity, and nuclear activity on star-formation rates based on H$\alpha$ emission}

\author[K. Kouroumpatzakis \etal]{K. Kouroumpatzakis$^{1,2}$\orcidA{}\thanks{E-mail: kkouroub@ia.forth.gr},
\author x A. Zezas$^{1,2,3}$\orcidB{},
\author x A. Maragkoudakis$^{4}$\orcidH{},
\author x S. P. Willner$^{3}$\orcidG{}, 
\newauthor
\author x P. Bonfini$^{1,2,5}$\orcidE{},
\author x M. L. N. Ashby$^{3}$\orcidF{},
\author x P. H. Sell$^{6}$\orcidC{},
and \author[K. Kouroumpatzakis \etal] f T. H. Jarrett$^{7}$\orcidI{}\\
$^{1}$Department of Physics, Univ. of Crete, GR-70013 Heraklion, Greece\\
$^{2}$Institute of Astrophysics, FORTH, GR-71110 Heraklion, Greece\\
$^{3}$Center for Astrophysics \textbar\ Harvard \& Smithsonian, 60 Garden St., Cambridge, MA 02138, USA\\
$^{4}$NASA Ames Research Center, Moffett Field, CA 94035-0001, USA\\
$^{5}$Institute for Astronomy, Astrophysics, Space Applications \& Remote Sensing, National Observatory of Athens, P. Penteli, GR-15236 Athens, Greece\\
$^{6}$Department of Astronomy, University of Florida, Gainesville, FL, 32611 USA\\
$^{7}$Department of Astronomy, University of Cape Town, Private Bag X3, Rondebosch 7701, South Africa\\
}

\date{Accepted for publication in MNRAS on 01/07/2021. Received 26/04/2021}

\pubyear{2020}

\begin{document}
\label{firstpage}
\pagerange{\pageref{firstpage}--\pageref{lastpage}}
\maketitle

\begin{abstract}

We present new H$\alpha$ photometry for the Star-Formation Reference Survey (SFRS), a representative sample of star-forming galaxies in the local Universe. 
Combining these data with the panchromatic coverage of the SFRS, we provide calibrations of H$\alpha$-based star-formation rates (SFRs) with and without correction for the contribution of [\ion{N}{II}] emission.
We consider the effect of extinction corrections based on the Balmer decrement, infrared excess (IRX), and spectral energy distribution (SED) fits. 
We compare the SFR estimates derived from SED fits, polycyclic aromatic hydrocarbons, hybrid indicators such as 24\,$\mu$m + H$\alpha$, 8\,$\mu$m + H$\alpha$, FIR + FUV, and \Ha\ emission for a sample of purely star-forming galaxies. 
We provide a new calibration for 1.4\,GHz-based SFRs by comparing to the H$\alpha$ emission, and we measure a dependence of the radio-to-H$\alpha$ emission ratio based on galaxy stellar mass. 
Active galactic nuclei introduce biases in the calibrations of different SFR indicators but have only a minimal effect on the inferred SFR densities from galaxy surveys.
Finally, we quantify the correlation between galaxy metallicity and extinction. 

\end{abstract}



\section{Introduction}
\label{sec:Introduction}

Star formation is one of the defining properties of galaxies. 
Since the reionization epoch, star formation has been transforming the primordial gas into stars, building what is now seen as the stellar mass ($M_\star$) of the galaxies while enriching the gas with metals.
The star-formation rate (SFR) of galaxies or large stellar populations (SPs) quantifies the recent or current star formation \citep[for a review see][]{2012ARA&A..50..531K}.
There are various indicators tracing SFR across the electromagnetic spectrum, based on emission produced by different physical mechanisms associated with star formation.
These mechanisms involve e.g., the direct ultraviolet (UV) emission from the photospheres of hot stars or the indirect emission by dust that has absorbed the UV radiation and re-emitted it in the infrared (IR).
However, the target of all SFR indicators is to measure the amount of massive/young stars residing in the SPs.
An assumption on the initial mass function \citep[IMF; e.g.,][]{1955ApJ...121..161S,1979ApJS...41..513M,2001MNRAS.322..231K,2003PASP..115..763C} can extrapolate this measurement and give an estimation of the total amount of all the recently born stars. 

Many studies of SFR indicators so far \citep[e.g.,][]{2006ApJS..164...81M,2007ApJ...666..870C,2008ApJS..178..247K,2009ApJ...703.1672K,2009ApJ...692..556R,2011ApJ...741..124H,2010A&A...518L..70B,2010ApJ...714.1256C,2011ApJ...737...67M,2017ApJ...850...68C,2017MNRAS.466.2312D,2019MNRAS.482..560M} have established a framework for measuring SFR.
However, these studies also showed that measuring star formation is a complex task, subject to systematic effects and often limited by the available data. 

For local-Universe galaxies, \Ha\ ($\lambda = 6563$~\AA) emission is one of the most widely used SFR indicators.
\Ha\ emission is produced when massive young stars ionize atomic hydrogen gas.
The electrons cascading through the atomic hydrogen energy levels emit radiation, producing the well studied \textit{Lyman}, \textit{Balmer}, \textit{Paschen} etc. lines.
The great abundance of hydrogen gas in all star-forming galaxies makes \Ha\ a strong emission line, easily detected when the galaxies host O stars that can ionise the interstellar medium (ISM) with their UV emission. 
The fact that the \Ha\ emission line is the strongest hydrogen recombination line in the visible range for the local Universe galaxies has made it one of the most commonly used tracers of star formation.

Moreover, \Ha\ is the ideal SFR indicator for probing stellar populations with ages less than 10~Myr \citep[e.g.,][]{1998ARA&A..36..189K,2012AJ....144....3L,2014A&A...571A..72B,2016A&A...589A.108C,2017ApJ...840...44B,2020MNRAS.498..235H}. 
In this sense, it is the closest indicator to the instantaneous SFR.
Tracing the youngest stellar populations, unbiased by emission arising from older stars, is important for studies related to these very short-lived stars, such as correlations with the X-ray emission from high-mass X-ray binaries \citep[HMXBs; e.g.,][]{2020MNRAS.494.5967K}.
This is also relevant for deriving scaling relations between stellar populations and their endpoints (neutron stars and stellar black holes) because the progenitors of these stellar remnants are very massive stars.

Despite its advantages, \Ha\ emission is subject to effects that can bias the SFR measurements. 
These effects include absorption by dust in the vicinity of the star-forming regions \citep[birth clouds; e.g.,][]{1994ApJ...429..582C,2000ApJ...539..718C} or the general ISM of a galaxy and the contribution of the adjacent-in-wavelength [\ion{N}{II}] $\lambda\lambda6548,~ 6583$ emission lines \citep[e.g.,][]{2008ApJS..178..247K,2021MNRAS.500..962K}.
Failing to correct for these effects or lack of required information (e.g., extinction measurements from independent methods) can lead to systematic differences from the actual SFR.

This work presents a systematic study of \Ha-based SFRs in comparison with other SFR indicators in a representative sample of nearby star-forming galaxies. 
Through complete photometric coverage of the sample from the UV to radio wavelengths, which also includes optical spectral information, we provide calibrations for the \Ha-based SFRs with respect to various extinction indicators and the [\ion{N}{II}] contribution.
We examine the SFRs derived by spectral energy distribution (SED) fits, 1.4~GHz radio emission, and hybrid indicators which combines 8 or 24~$\mu$m and \Ha\ emission \citep[e.g.,][]{2007ApJ...666..870C}.
Furthermore, we explore the relation between 1.4~GHz emission and SFR and the effect of active galactic nuclei (AGN) in the SFR measurements and calibrations.
Finally, we examine the connection between  extinction and metallicity.

Section \ref{sec:Observations} describes the sample and basic data and Section \ref{sec:Ha_photometry} the \Ha\ photometry.
Section \ref{sec:Results} presents the 
calculation of \Ha\ SFRs and the relevant corrections, as well as comparisons with other SFR indicators for star-forming galaxies.
Sections \ref{sec:Discussion} and \ref{sec:Summary} discuss and summarize the results of this work.
All linear regressions were robust linear-regression fits performed with the \texttt{Python statsmodel RLM} package \citep[][]{seabold2010statsmodels}, 
which is designed to be uninfluenced by outliers.
All reported uncertainties are at the 68\% confidence level.
We assume a cosmology with $\Omega_m=0.3$, $\Omega_\Lambda=0.7$, $h=0.73$ consistent with \cite{2011PASP..123.1011A}. 
We adopt as solar abundances $\rm Z_{\odot} = 0.0142$, $\rm X_{\odot} = 0.7154$, and [12 + log(O/H)$_\odot$] = 8.69 \citep{doi:10.1146/annurev.astro.46.060407.145222}.

\section{Sample and Observations}
\label{sec:Observations}

\subsection{Sample}
\label{sec:Sample}

The basis of our sample is the Star Formation Reference Survey \citep[SFRS;][]{2011PASP..123.1011A}. 
SFRS comprises 369 local-Universe galaxies (${\langle} z {\rangle} = 0.018, z_{\rm max} = 0.30641$) selected to represent star formation under various conditions present in the local Universe.
More specifically, SFRS was selected out of the parent PSC$z$ catalog \citep{2000MNRAS.317...55S} to cover the 3D space of the fundamental galactic properties of SFR, specific SFR (sSFR), and dust temperature. 
In this selection, SFR was indicated by the $L_{\rm 60\,\mu m}$, sSFR by the  $f_{\rm K_S}-f_{\rm 60 \mu m}$\,colour, and dust temperature by the $f_{\rm 100\,\mu m}/f_{\rm 60\,\mu m}$ flux ratio.
SFRS galaxies have wide coverage of the electromagnetic spectrum, from radio to X-rays \citep{2019MNRAS.482..560M,2020MNRAS.494.5967K}, including optical spectra of the galaxy nuclei \citep{2018MNRAS.475.1485M} as well as parameterization of their morphology including decomposed bulge and disk stellar masses \citep[][]{2021MNRAS.504.3831B}.

The SFRS galaxies where classified in four classes: star-forming, Seyfert, transition objects (TO; a.k.a. composite), and LINER \citep[][see their Figure 7]{2018MNRAS.475.1485M} based on an ionization source classification through emission-line ratio diagrams \citep[BPT diagrams;][]{1981PASP...93....5B,1987ApJS...63..295V,2001ApJ...556..121K,2003MNRAS.346.1055K,2007MNRAS.382.1415S}.
In the following analysis, we consider only purely star-forming (not hosting AGN) galaxies as characterized by \cite{2018MNRAS.475.1485M}, except in Section \ref{sec:Disc+AGN}, where we discuss the effect of AGN in SFR measurements and calibrations.
Therefore, the ionization in these galaxies is driven by star formation rather than being dominated by other sources such as shocks or AGN.

\begin{figure}
    \centering
    \includegraphics[width=\linewidth]{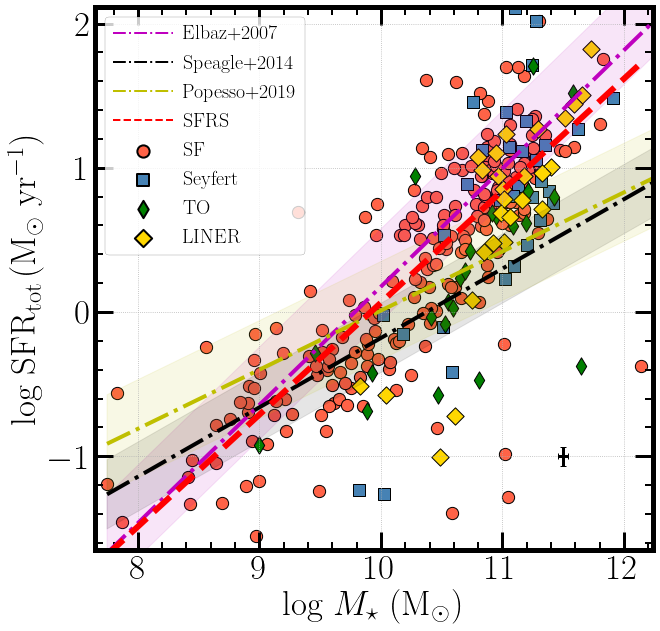}
    \caption{SFR as a function of stellar mass for the SFRS.  
    Galaxies classified as star-forming, Seyfert, TOs, and LINER are shown with red circles, blues squares, green rhombus, and yellow diamonds respectively.
    The black error bar at the bottom right indicates the median uncertainties.
    Inferred SFRs for non star-forming galaxies are only indicative because we do not account for AGN or post-AGB stars contribution.
    The red dashed line represents the best fit for the SFRS star-forming galaxies (Section \ref{sec:Sample}).
    The magenta, black, and yellow dashed-dotted lines represent the MS for diverse samples of local Universe galaxies from
    \protect\cite{2007A&A...468...33E}, \protect\cite{2014ApJS..214...15S}, and \protect\cite{2019MNRAS.483.3213P} respectively.}
    \label{fig:MS}
\end{figure}

Figure \ref{fig:MS} shows the \textit{Main Sequence of star-forming galaxies} (MS) for the SFRS.
The best fit for the SFRS star-forming galaxies is:
\begin{eqnarray}
    {\rm log ~\frac{SFR}{(M_\odot ~ yr^{-1})}} = -7.7 (\pm 0.32) + 0.78 (\pm 0.03) ~ {\rm log}~\frac{M_\star}{(\rm M_\odot)} \quad .
\end{eqnarray}
The MS relation for the SFRS sample is consistent with the one reported by \cite{2017MNRAS.466.1192M}, which was based on the \textit{Spitzer} $8~\mu$m emission SFRs.
However, here we calculate it using our reference SFR indicator ($\rm SFR_{tot}$; see Section \ref{sec:SFR}).
The SFRS covers a great range below and above the MS reflecting the wide range of $f_{\rm K_S}-f_{\rm 60 \mu m}$ indices (a proxy for sSFR) used to construct the sample.
Figure \ref{fig:MS} shows all classes of galaxies discussed here. 
However, the inferred SFRs for non star-forming galaxies are only indicative because we do not account for the possible contamination from AGN or post-AGB stars.
The SFRS MS is in good agreement with other samples, considering that the slope of the MS can vary significantly depending on the SFR indicator and sample used.
There is particularly good agreement with the MS from \cite{2007A&A...468...33E} which, like SFRS, is also based on an IR-selected sample and IR-based SFRs (in contrast the \cite{2014ApJS..214...15S}, and \cite{2019MNRAS.483.3213P} works are based on diverse samples and SFR indicators).

\subsection{\texorpdfstring{H$\alpha$}~~observations}

\begin{figure}
    \centering
    \includegraphics[width=\linewidth]{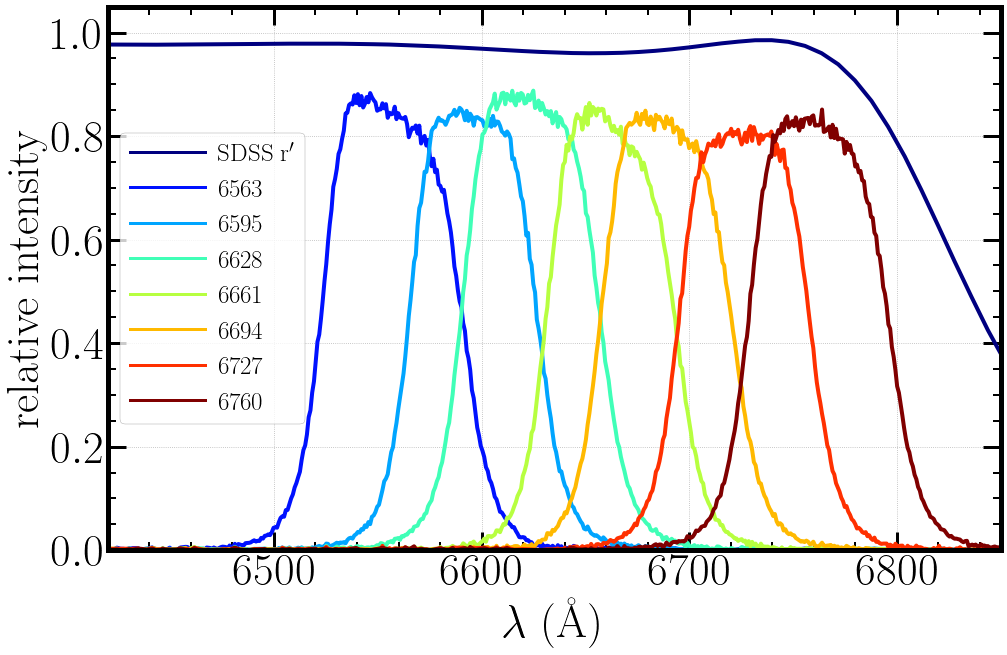}
    \caption{Transmission curves of the filters used in the \Ha\ imaging campaign as measured for the $f{/}7.6$ focal ratio of the Skinakas telescope.
    The $f{/}8$ FLWO telescope will have nearly identical transmissions.
    The central wavelength of each filters is indicated in the legend with the same color as in the graph.}
    \label{fig:Filters}
\end{figure}

\begin{figure}
    \centering
    \includegraphics[width=\linewidth]{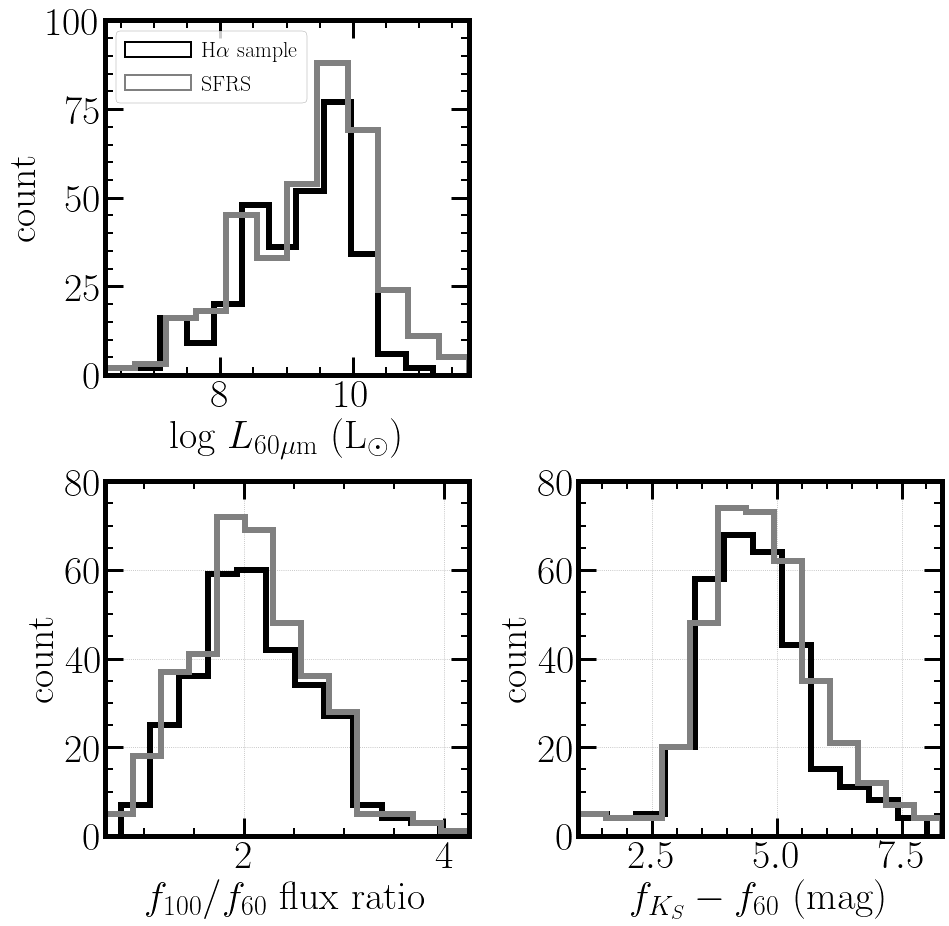}
    \caption{Comparison of the SFRS selection properties: $L_{\rm 60 \mu m}$ (top left; a proxy of SFR), $f_{\rm 100\mu m}/f_{\rm 60\mu m}$ (bottom left; a proxy of dust temperature), and $f_{K_{\rm S}} - f_{\rm 60\mu m}$ (bottom right; a proxy of sSFR), between the entire SFRS (gray), and the sample with \Ha\ observations (black).}
    \label{fig:SFRS_Ha_sample}
\end{figure}

We obtained \Ha\ + [\ion{N}{II}] and nearby continuum imaging observations for 305 SFRS galaxies with the 1.3~m telescope of the Skinakas\footnote{\url{http://skinakas.physics.uoc.gr/en/}} observatory and the 1.2~m telescope of the Fred Lawrence Whipple observatory\footnote{\url{http://www.sao.arizona.edu/}} (FLWO) for a total of ${\sim} 180$ nights at Skinakas and 31 nights at FLWO.
We used a custom set of narrow-band filters to adjust for the redshift range of the SFRS galaxies, centered at $\lambda
=$ 6563, 6595, 6628, 6661, 6694, 6727, and 6760~\AA\ with typical $\rm FWHM=65$\,\AA\ and a filter equivalent to SDSS~$r'$ for the continuum observations (Figure \ref{fig:Filters}). 
SFRS galaxies with $z > 0.03$ could not be covered by the available filters.
Therefore, these higher-redshift SFRS galaxies were not observed, resulting in a total of 305 out of 369 galaxies with \Ha\ observations.
Because the selection function of the SFRS is highly insensitive to redshift, this limitation did not affect the following analysis except for slightly limiting the high end of SFR values (Figure \ref{fig:SFRS_Ha_sample}).
Among the galaxies excluded were the luminous AGNs 3C~273 and OJ~287.

The \Ha\ and $r'$ observations received exposure of 1\,hour and ${\sim}$10 minutes respectively. 
The observations took place under photometric conditions and typical seeing FWHM ${\sim} 1.2\arcsec$ and ${\sim} 1.7\arcsec$ for Skinakas and FLWO respectively.
Some galaxies were re-observed when in doubt about the photometric conditions of the original observations.
The total exposure of each galaxy observed in \Ha\ was split in either 6 observations of 600~seconds or 12 observations of 300~seconds.
Thus, the subtraction of cosmic rays was more efficient, and photometric-conditions variations were monitored during the observations.

\section{\texorpdfstring{H$\alpha$}~~photometry}
\label{sec:Ha_photometry}

\subsection{Basic reduction and continuum subtraction}
\label{sec:Basic_reduction}

The basic reduction was performed with \texttt{IRAF} \citep{1986SPIE..627..733T,1993ASPC...52..173T}.
The task \texttt{ccdproc} was used for the bias subtraction and flat fielding. 
The separate frames were aligned and combined with the \texttt{imalign} and \texttt{imcombine} tasks respectively. 
Astrometry was applied to the final combined frames with \textit{Astrometry.net} \citep{2010AJ....139.1782L}.

\begin{figure}
    \centering
    \hbox{\includegraphics[width=0.5\linewidth]{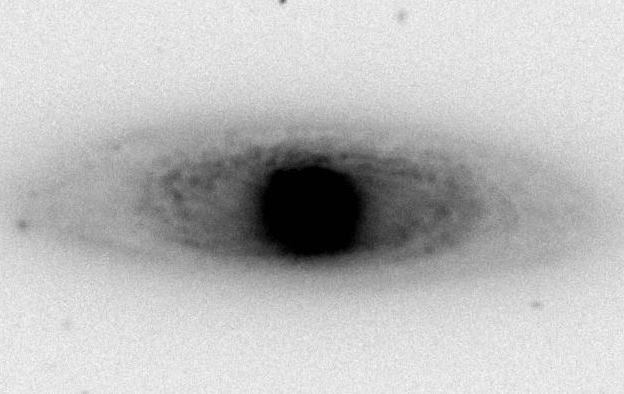}
    \includegraphics[width=0.5\linewidth]{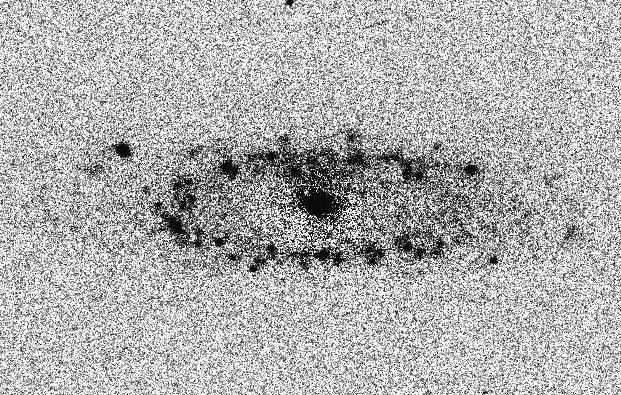}}
    \caption{Example of continuum subtraction in galaxy NGC\,4448.
    In the left plot is the observed \Ha, and in the right is the continuum-subtracted \Ha\ image.}
    \label{fig:Cont_sub_image}
\end{figure}

In order to perform continuum subtraction to the \Ha\ images, we followed the standard procedures for narrow-band imaging \citep[e.g.,][]{2008ApJS..178..247K}.
We first measured the flux of the foreground stars in both the \Ha\ and continuum red images using the \texttt{IRAF} task \texttt{daophot}.
The mode and the standard deviation of the \Ha\ to $r'$ continuum flux ratio distribution were used as the continuum-subtraction ratio $F$ and its uncertainty respectively.
From the  \Ha\ image, we subtracted the continuum image scaled by $F$ in order to produce the final continuum-free \Ha\ image (e.g., Figure  \ref{fig:Cont_sub_image}):

\begin{eqnarray}
    \rm Image_{H\alpha \, cont. \,sub.} = Image_{H\alpha} - Image_{cont. \, r'} \times  \textit{F} \quad .
\end{eqnarray}

\begin{figure}
\begin{center}
    \includegraphics[width=0.8\linewidth]{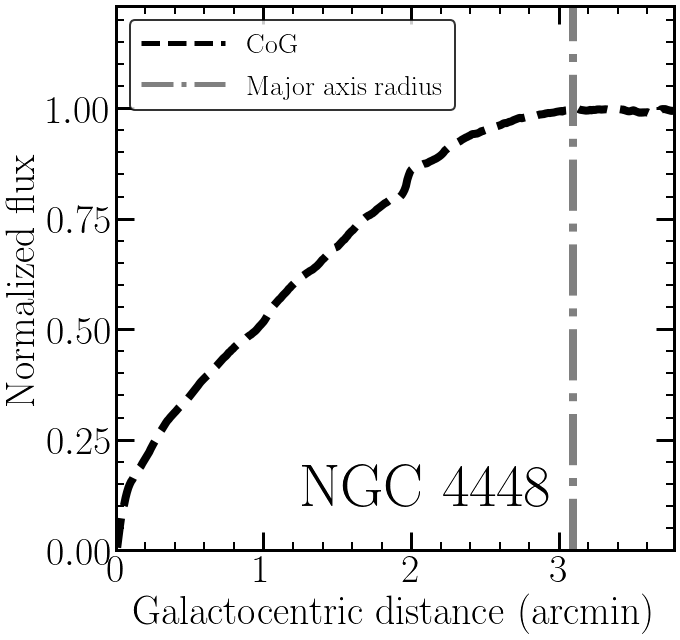}
    \caption{
    Curve of growth (dashed black line) for the H$\alpha + [\ion{N}{II}]$ flux of galaxy NGC\,4448. 
    The normalized integrated flux is shown on the vertical axis as a function of the aperture semimajor axis. 
    The semimajor axis of the adopted aperture is shown by a dashed-dotted grey line.}
    \label{fig:CoG}
\end{center}
\end{figure}

As described by \cite{2020MNRAS.494.5967K}, the curve of growth (CoG) technique was used on the continuum-subtracted images to measure the flux of the galaxies. The CoG technique has the benefit of determining the optimal aperture that contains the total flux of the object while  also measuring and subtracting the background contribution.
The shape of the apertures was based on elliptical-aperture fits to WISE 4.6~$\mu$m data of the SFRS galaxies following a procedure similar to \cite{2019ApJS..245...25J}.
The CoG was calculated by increasing the aperture radius up to a maximum radius $r_{\rm max}$, while keeping the position angle of the ellipse and the major-to-minor axis ratios fixed. 
The maximum radius of the CoG $r_{\rm max}$ was required to be beyond the D25 isophote in order to encompass the total galaxy emission. 
This was confirmed by visually inspecting the results of the process.
The resolution of the CoG ranged from 1 to 5 pixels for small to large (in aperture size) galaxies, respectively.
In order to measure the asymptotic line of the CoG, a linear-regression fit was performed to the last 5\% of the CoG points.
The CoG was iteratively repeated, while adjusting the background, until the regression-fit slope (of the last 5\% of the CoG) was zero.
Then, the size of the aperture was defined by the smallest radius of the CoG that reached the asymptotic line (example in Figure ~\ref{fig:CoG}).

\subsection{Photometry }
\label{sec:photometry}

In order to account for differences in the filter transmission curves between those measured with a parallel beam (usually reported by filter manufacturers) and the telescope's conical beam, we measured the filters' transmission curves with a lens of the same focal ratio as the Skinakas 1.3m telescope ($f = 7.6$; Figure \ref{fig:Filters}).

In order to account for photometric variations during each observation, we incorporated the standard deviation of the distribution of fluxes of all the observing frames in the calculation of the photometric uncertainty.
For the photometric absolute calibrations, we observed spectrophotometric standard stars \citep{1988ApJ...328..315M} at various airmasses during each observing run.
We followed the standard procedure of fitting the instrumental magnitude as function of the airmass $\chi$ in order to measure the zero-point (ZP) and the atmospheric attenuation $\kappa$ for each night. 
The reference magnitude of standard stars at the top of the atmosphere was calculated by integrating the reference spectrum $S$ with the filter response $R$:

\begin{eqnarray}
    m_{\rm ref} = -2.5 \, {\rm log} \Bigg( \frac{\int{(R \times S) ~ d\lambda}}{\int{S} ~ d\lambda} \Bigg) \quad .
\end{eqnarray}
The ZP was then calculated as:
\begin{eqnarray}
    \rm ZP = \textit{m}_{ref} + 2.5 \, log(CR) \quad ,
\end{eqnarray}
where CR is the count rate of the observed standard star.

The CRs from the integrated photometry of each object were converted to flux following the scheme presented by \cite{2008ApJS..178..247K}.
We have also included in our calculation the transmission correction \citep[see Appendix A in][]{2008ApJS..178..247K} that accounts for the differential transmission of the narrow and broad band filters.
This correction takes into account the position of the \Ha\ and [\ion{N}{II}] emission lines with respect
to the transmission curve of the filter and the redshift of each galaxy.
Therefore, we calculated the flux $f_\lambda$ using:

\begin{eqnarray}
    f_\lambda = \lambda^2 \, 10^{-0.4 \, [{\rm ZP} + 2.397 - \kappa {\rm sec}(x)]} \, T_{\rm c} \quad ,
\end{eqnarray}
\begin{eqnarray}
T_{\rm c} = {\rm FWHM_{NB} \, CR} \, \Bigg[ T_{\rm NB}(\lambda) - T_{\rm R}(\lambda) \frac{t_{\rm R}}{t_{\rm NB}}\frac{1}{F} \Bigg]^{-1} \quad ,
\end{eqnarray}
where $\rm FWHM_{NB}$ is the full-width-half-maximum of the narrow-band filter, CR is the count-rate [counts $\rm s^{-1}$], $T_{\rm R}$ and $T_{\rm NB}$ are the normalized filter transmissions, $t_{\rm R}$ and $t_{\rm NB}$ are the exposure times of the continuum and the narrow-band filter respectively, and $F$ is the continuum-subtraction ratio.

The normalized transmission of the filters ($T_{\rm NB}$ and $T_{\rm R}$) accounts for the different transmission at the wavelengths of the \Ha\ and [\ion{N}{II}] emission lines by their respective transmissions at the wavelengths of the \Ha\ and [\ion{N}{II}] lines (at the observed frame) weighted by the average \Ha\ to [\ion{N}{II}] ratio.
We ignored the contribution of the [\ion{N}{II}]$\lambda6548$ line in the calculation of the filter transmission because of its significantly smaller intensity in comparison to the \Ha\ and the [\ion{N}{II}]$\lambda6584$ lines. 
The contribution of the [\ion{N}{II}] flux was subtracted in the next step of the analysis (see Section \ref{sec:Discussion_Ha_NII}). 
The Appendix presents the H$\alpha$~+~[\ion{N}{II}] fluxes and luminosities for the SFRS galaxies using distances from \cite{2011PASP..123.1011A}.

\subsection{\texorpdfstring{H$\alpha$}~ photometry comparison}
\label{sec:Photometry_comparison}

\begin{figure}
    \centering
    \includegraphics[width=\columnwidth]{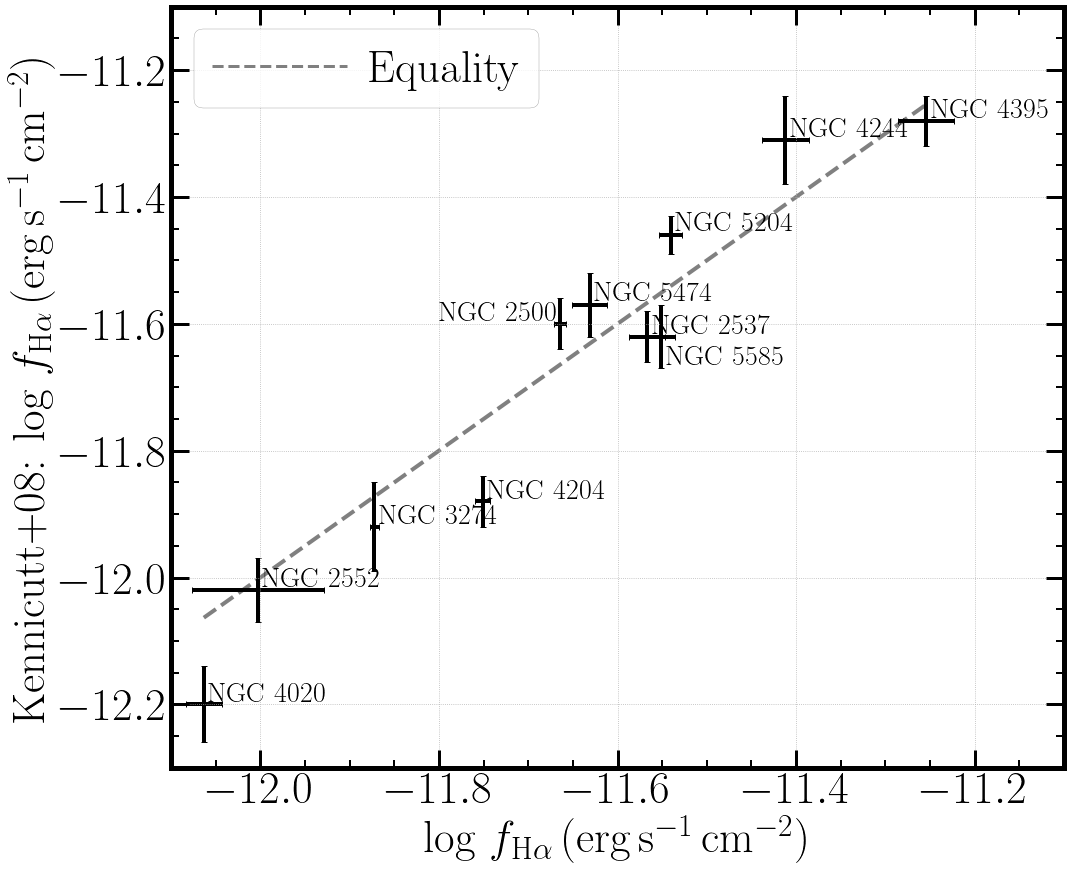}
    \caption{Comparison of the logarithm of the \Ha\ + [\ion{N}{II}] flux measurements between \protect\cite{2008ApJS..178..247K} (ordinate) and this work (abscissa).
    Dashed line shows equality.}
    \label{fig:ken_comp}
\end{figure}

Figure \ref{fig:ken_comp} compares our \Ha\ photometry with that of the \Ha\ survey of nearby galaxies within 11~Mpc by \cite{2008ApJS..178..247K}.
The comparison for the 11 galaxies in common shows excellent agreement between the flux measurements of the two surveys.
The standard deviation of the ratio between the two surveys is $\delta ({\rm log}~{f_{\rm H\alpha}^{\rm K21}}/{f_{\rm H\alpha}^{\rm K08}}) = 0.08$~dex, where $\rm K21$ indicates this work.

\section{Data and results}
\label{sec:Results}

\subsection{Star-formation rates}
\label{sec:SFR}

The H$\alpha$ and H$\alpha$~+~[\ion{N}{II}] luminosities were converted to SFR through the theoretical \cite{2011ApJ...737...67M} relation:
\begin{eqnarray}
   \frac{ {\rm SFR_{H\alpha}}}{ {\rm (\msun\,  yr^{-1})}} = 5.37 ~ \frac{ {L}_{\rm H\alpha}  }{{\rm ({10^{42} ~ erg\,  s^{-1}})}}  \quad .
	\label{eq:SFR-Ha}
\end{eqnarray}

In addition to the \Ha\ emission, we have also used in this work SFRs based on several different continuum bands, as well as composite indicators resulting from combinations of these bands. 
These were presented by \cite{2019MNRAS.482..560M}. 
The SFRs reported there include the radio 1.4\,GHz emission, tracing synchrotron radiation from 
relativistic electrons produced in supernovae;
FIR emission from dust heated by UV emission from young stars; 8$\mu$m emission from polycyclic aromatic hydrocarbons (PAHs), tracing the photo-dissociation regions around young stellar populations; and UV emission from the photospheres of OB stars. 
We also used the combination of the $\rm SFR_{FUV}$ and $\rm SFR_{FIR}$:
\begin{eqnarray}
    \rm SFR_{tot} = SFR_{FUV} + (1 - \eta) \, SFR_{FIR} \quad ,
\end{eqnarray}
where $\eta$ refers to the fraction of FIR emission that is associated with old stars rather than the dust heated by the massive/young stars.
\cite{2019MNRAS.482..560M} adopted $\eta = 0.0$ as the nominal value for the SFRS sample, based on comparisons of the FIR and TIR emission, showing that the FIR is a SFR indicator, less biased by the thermal stellar emission, with respect to the TIR.
This was supported by the fact that the correlation of $\rm SFR_{tot}$ with $\rm SFR_{1.4~GHz}$ shows a preference to $\eta = 0.0$ (See also Section \ref{sec:Discussion_SFR_radio}).
The uncertainty of $\rm SFR_{tot}$ is dominated by the uncertainties of the IRAS fluxes, which are on average 10--20\% \citep[][]{2011PASP..123.1011A}.
For this work, we adopt an average 15\% uncertainty for the $\rm SFR_{tot}$.

The 8~$\mu$m emission is produced in the photo-dissociation regions of star-forming bubbles from PAH molecules excited by UV radiation. 
In order to account for the 8\,$\mu$m emission attributed to PAHs, the stellar continuum was estimated and subtracted using the formula from \cite{2007ApJ...664..840H}:
\begin{eqnarray}
    {f_{8 \micron,{\rm{PAH}}} = f_{8 \micron} - 0.227 \, f_{3.6 \micron}}\quad.
	\label{eq:Helou} 
\end{eqnarray}
Then PAH 8\,$\mu$m luminosity was converted to SFR with the calibration of \cite{2005ApJ...632L..79W}:
\begin{eqnarray}
    \frac{{\rm SFR}_{8\micron,{\rm PAH}}}{(\msunpyr)} = 
    \frac{\nu L_{8 \micron,{\rm{PAH}}}}{(1.57 \, 10^{9} ~ {\rm L_\odot})}\quad.
	\label{eq:SFR_8}
\end{eqnarray}

We have also calculated SFRs based on the 24~$\mu$m emission, which is produced by the UV-heated dust, using the 24~$\mu$m fluxes reported by \cite{2011PASP..123.1011A}.
The 24~$\mu$m luminosities were converted to SFR using the calibration of \cite{2009ApJ...692..556R}:
\begin{eqnarray}
    \frac{\rm SFR_{24\micron}}{(\msunpyr)} = 2.04 \,
    \frac{\nu {L}_{24\micron}}{ (10^{43}~\erg) }\quad,
	\label{eq:SFR-24}
\end{eqnarray}
where $\nu$ is the effective frequency of the band.
The 24~$\mu$m flux uncertainties are dominated by the uncertainty in the absolute calibration, which is estimated as 4--8\% according to the MIPS \textit{Instrument Handbook}. 
For this work, we adopt an average 6\% uncertainty for the $\rm SFR_{24\micron}$.

We also considered the WISE band-3 (12~$\mu$m) and band-4 (23~$\mu$m) luminosities, which were converted to SFR using the
calibrations of \cite{2017ApJ...850...68C}:
\begin{eqnarray}
    {\rm log}~\frac{\rm SFR_{W3}}{(\msunpyr)} = 0.889~
    {\rm log}~\frac{{L}_{W3}}{ (L_\odot) } - 7.76 \quad ,
	\label{eq:SFR-WISE3}
\end{eqnarray}
and 
\begin{eqnarray}
    {\rm log}~\frac{\rm SFR_{W4}}{(\msunpyr)} = 0.915~
    {\rm log}~\frac{{L}_{W4}}{ (L_\odot) } - 8.01 \quad .
	\label{eq:SFR-WISE4}
\end{eqnarray}
A comparison of different SFR conversions involving WISE bands is presented in Section \ref{sec:Discussion_SFR_WISE}.

The \cite{2019MNRAS.482..560M} SFRs were based on a Salpeter IMF \citep{1955ApJ...121..161S}.
In order to compare them with the \Ha\ and 24~$\mu$m-based SFRs which were based on a Kroupa IMF \citep[][]{2001MNRAS.322..231K}, all SFRs derived by \cite{2019MNRAS.482..560M} were multiplied with a scaling factor of $\rm SFR_{Kroupa}/SFR_{Salpeter}=0.67$ \citep{2014ARA&A..52..415M}.

A widely used and robust method for estimating galaxy properties when multi-band photometry is available comes from the analysis of their SEDs.
The SFRS galaxies benefit from such multi-wavelength coverage.
SFR estimation from SED fits for the SFRS galaxies are presented using the \textsc{cigale} \citep{2005MNRAS.360.1413B,2009A&A...507.1793N,2019A&A...622A.103B} SED-fitting code.
The basis of the \textsc{cigale} code is the energy balance between the dust-absorbed energy with the energy that is re-emitted in the MIR and FIR.
The galaxies were fitted assuming a delayed star-formation history with an additional exponential burst for modeling the latest star formation episode. 
The low-resolution \cite{BC03} single-stellar-population library was adopted to compute the intrinsic stellar spectrum. 
Nebular emission was modeled using \textsc{cigale}'s default \texttt{nebular} module, which is based on the \cite{Inoue2011} nebular templates generated with \textsc{cloudy} \citep{Ferland1998, Ferland2013}. 
The \cite{2000ApJ...539..718C} attenuation model was employed in order to account for the differential attenuation between young and old stars, with the former being subjected to an additional attenuation component of their birth clouds (BC), along with the attenuation from the ISM being common for both old and young stellar populations. 
The \cite{DL07} dust emission models were used to model the re-emission in the mid- and far-IR of the absorbed UV photons. 
Galaxy redshifts were provided as input and were not modeled in \textsc{cigale}. 
183 star-forming SFRS galaxies were successfully modeled with \textsc{cigale} with reduced $\chi^{2} < 5$.
More details about the SED fits of the entire sample of SFRS galaxies will be presented by Maragoudakis~et~al.(in prep.).
The SED-based SFRs are reported in the Appendix.

\subsection{Extinction indicators}
\label{sec:Extinction_indicators}

Visible and UV emission can be partially or completely absorbed by dust.
Therefore, in order to correctly estimate the SFRs from the \Ha\ emission, we need to know the extinction.
There are multiple ways to estimate the extinction depending on the available data.
This work compares extinction measurements from the Balmer decrement (based on the $f_{\rm H\alpha} / f_{\rm  H\beta}$ ratio), the IR excess (IRX; based on  $f_{\rm FIR}/f_{\rm FUV}$ ratio; e.g., \citealt{1996A&A...306...61B,1999ApJ...521...64M,2000ApJ...533..236G}), and  SED fits.
\cite{2018MNRAS.475.1485M} measured the flux of the \Ha\ and \Hb\ lines, extracted from the nuclear regions of the SFRS galaxies. 
The same work also gave, for the subset of the sample with long-slit spectra, Balmer line measurements from larger apertures extending over the major axis of the galaxies.

\begin{figure}
    \centering
    \includegraphics[width=\columnwidth]{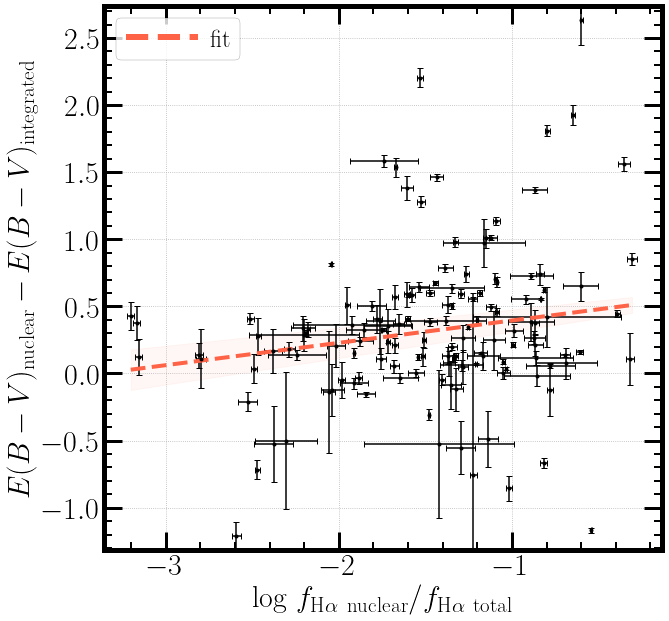}
    \caption{Comparison of $E(B{-}V)$ values based on the Balmer decrement derived from nuclear and wide spectral extraction apertures as a function of the ratio of the \Ha\ flux measured in the same apertures.
    The best-fit linear relation (slope $a = 0.17 \pm 0.06$ and intercept $0.56 \pm 0.1$) is presented with a red dashed line.
    }
    \label{fig:EBV_Ha_nuc_tot}
\end{figure}

The Balmer extinction was estimated through the conversion of \cite{2013ApJ...763..145D}:
\begin{eqnarray}
    E(B{-}V) = 1.97 \, {\rm log} \left[\frac{(f_{\rm H\alpha} / f_{\rm  H\beta} )}{2.86}\right] \quad ,
\end{eqnarray}
using the reddening law of \cite{2000ApJ...533..682C}.
We adopted $E(B{-}V) = 0$ for 18 galaxies with $f_{\rm H\alpha}/f_{\rm H\beta} < 2.86$.

We adopted the integrated-galaxy-emission Balmer decrements for 92 galaxies with such available data. 
For the rest of the galaxies where only nuclear-aperture spectra were available, a corrected Balmer decrement $E(B{-}V)$ was calculated based on the relation between the difference of the ratio of the flux measured within the nuclear and integrated extraction apertures (Figure \ref{fig:EBV_Ha_nuc_tot}).
This way we corrected for the increased extinction in the nuclear regions of galaxies in comparison to the extinction from the long-slit apertures that sampled larger regions of the galaxies.

We adopted the IRX-based extinctions from \cite{2019MNRAS.482..560M}, which were calculated based on the integrated emission GALEX-FUV \citep[][]{2005ApJ...619L...1M} and FIR fluxes, and the $E(B{-}V)_{\rm IRX}$ calibration of \cite{2005ApJ...619L..51B}:

\begin{eqnarray}
    {\rm \textit{A}_{FUV}(IRX)} = -0.0333p^3 + 0.3522p^2 + 1.1960p + 0.4967
\end{eqnarray}
where $p = {\rm log} (L_{\rm FIR} / L_{\rm FUV})$, and
\begin{eqnarray}
    E(B{-}V)_{\rm IRX} = \textit{A}_{\rm FUV}/\textit{k}_{\rm FUV}  \quad ,
\end{eqnarray}
where $k_{\rm FUV} = 10.22$ from the \cite{2000ApJ...533..682C} attenuation law.

Extinction from SED fits was modeled after the \texttt{dustatt\_modified\_CF00} implemented in the \textsc{cigale} code and the \cite{2000ApJ...539..718C} attenuation model. 
This model includes attenuation for very young stars which are still embedded in their birth clouds (BCs) in addition to the attenuation for stars due to the general ISM. 
The attenuation curves attributed to each component are of the form $\rm A(\lambda) \propto \lambda^{\delta}$, where $\rm \delta_{BC} = -1.3$ and $\rm \delta_{ISM} = -0.7$.
Stars younger than 10~Myr are attenuated from both BCs and ISM components, while older stars are attenuated only from the ISM curve.
The three independent estimates of the extinction are reported in the Appendix.

\subsection{Comparison between extinction indicators}
\label{sec:Ext_comp}

\begin{figure*}
    \centering
    \includegraphics[width=0.8\textwidth]{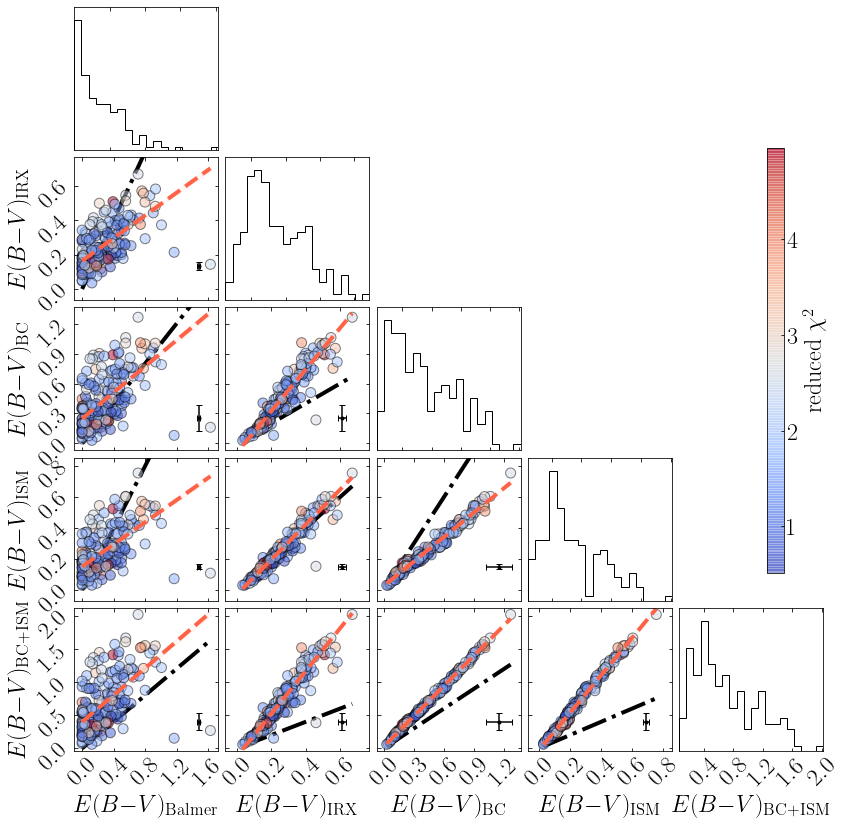}
    \caption{Comparisons between extinctions derived through the Balmer decrement, the IRX, and SED fits separated in the BC and ISM components and combined. 
    The black dashed-dotted lines show equality.
    The red dashed lines represent the linear-regression fits (Table \ref{tab:extinction}).
    Points are colour-coded based on the reduced $\chi^2$ of the SED fits.
    The black error bar at the bottom right of each panel indicates the median uncertainties of the respective extinction indicators.}
    \label{fig:corner_ext}
\end{figure*}

Figure \ref{fig:corner_ext} compares the extinctions derived from SED fitting, the Balmer decrement, and IRX (Section \ref{sec:Extinction_indicators}).
We quantify these comparison by their median difference, scatter and linear-regression fits (Table \ref{tab:extinction}).
This comparison involves the 183 star-forming galaxies with SED fits with reduced $\chi^2 < 5$, and we considered in addition to the different extinction indicators, the extinction from the BC and ISM components from the SED fits extinction (as inferred from the \texttt{dustatt\_modified\_CF00} mode; Section \ref{sec:Extinction_indicators}).

IRX-based extinction is slightly underestimated in comparison with the Balmer-decrement-based extinction in the high extinction regime. 
The IRX is in excellent agreement with the SED-fits ISM component extinction.
The birth-clouds component extinction is closer to the Balmer-decrement extinction, but the first tends to overestimate extinction almost in all cases.
The combination of the ISM and the birth-clouds components ($E(B{-}V)_{\rm ISM} + E(B{-}V)_{\rm BC}$) shows a linear relation with the Balmer decrement, but it tends to overestimate the extinction by $\sim 0.4$ mag.

\begin{table*}
    \centering
    \caption{Comparisons and calibrations between the Balmer, IRX and the separate ISM and birth-clouds (BC) components of the SED (\texttt{dustatt\_modified\_CF00}) extinction indicators.
    Linear-regression fits are of the form $E(B{-}V)_{\upsilon} = \alpha + \beta ~ E(B{-}V)_{\chi}$.}
    \begin{tabular}{cccccc}
    $\upsilon/\chi$ & ${\langle} E(B{-}V)_{\upsilon} - E(B{-}V)_{\chi} {\rangle} $ & $\delta {\langle} E(B{-}V)_{\upsilon} - E(B{-}V)_{\chi} {\rangle} $ & $\alpha$ & $\beta$\\
    \hline
    IRX/Balmer & $0.02$ & $0.22$ & $0.16 \pm 0.01$ & $0.33 \pm 0.03$\\
    BC/Balmer & $0.14$ & $0.27$ & $0.25 \pm 0.02$ & $0.65 \pm 0.06$\\
    BC/IRX & $0.15$ & $0.16$ & $-0.10 \pm 0.01$ &  $2.09 \pm 0.04$\\
    ISM/Balmer & $0.03$ & $0.23$ & $0.15 \pm 0.01$ &  $0.36 \pm 0.03$\\
    ISM/IRX & $0.00$ & $0.05$ & $-0.03 \pm 0.01$ & $1.13 \pm 0.02$\\
    ISM/BC & $-0.16$ & $0.13$ & $0.02 \pm 0.00$ & $0.53 \pm 0.01$\\
    (BC+ISM)/Balmer & $0.37$ & $0.37$ & $0.40 \pm 0.04$ & $1.01 \pm 0.09$\\
    (BC+ISM)/IRX & $0.36$ & $0.30$ & $-0.13 \pm 0.02$ & $3.23 \pm 0.06$\\
    (BC+ISM)/BC & $0.21$ & $0.15$ & $0.02 \pm 0.00$ & $1.53 \pm 0.01$\\
    (BC+ISM)/ISM & $0.38$ & $0.27$ & $-0.03 \pm 0.01$ & $2.83 \pm 0.03$\\
    \end{tabular}
    \label{tab:extinction}
\end{table*}

\subsection{\texorpdfstring{H$\alpha$}~~SFRs corrected for the \texorpdfstring{[\ion{N}{II}}\ ] contribution}
\label{sec:Ha_NII_contribution}

\begin{figure}
    \centering
    \includegraphics[width=\linewidth]{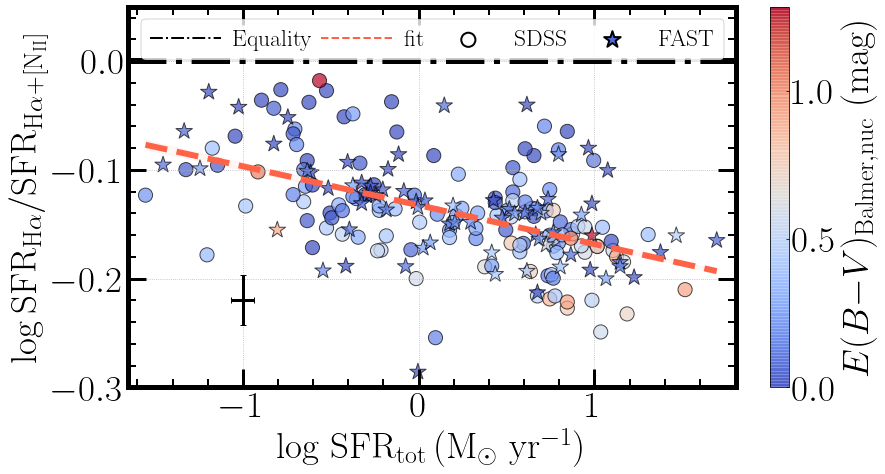}
    \caption{The ratio of SFRs derived from the \Ha\ luminosity corrected for the [\ion{N}{II}] contribution, over the un-corrected \Ha~+~[\ion{N}{II}], as a function of $\rm SFR_{tot}$.
    Circle and star markers show galaxies with spectra obtained with the SDSS (nuclear) or FAST (long-slit) spectrographs respectively.
    The points are colour-coded according to their Balmer-decrement-based $E(B{-}V)$.
    The red dashed line represents the linear-regression fit (reported in Section \ref{sec:Ha_NII_contribution}).
    The black error bar at the bottom left indicates the median uncertainties.}
    \label{fig:Ha_HaNII}
\end{figure}

The [\ion{N}{II}] flux contribution in the \Ha\ measurements was corrected based on the following relation:
\begin{eqnarray}
    f_{\rm H\alpha} = \frac{f_{\rm H\alpha+[\ion{N}{II}]}}{1 +  f_{\rm [\ion{N}{II}]}/f_{\rm H\alpha}} \quad ,
\end{eqnarray}
where $f_{\rm H\alpha+[\ion{N}{II}]}$ is the observed photometric flux, and $f_{\rm [\ion{N}{II}]}/f_{\rm H\alpha}$ is the flux ratio between the [\ion{N}{II}] and H$\alpha$ emission as measured from spectroscopic observations.
Figure \ref{fig:Ha_HaNII} presents the \Ha-derived SFRs corrected for the contribution of the adjacent [\ion{N}{II}] emission lines, compared to un-corrected ones, for 260 star-forming SFRS galaxies. 
The fact that the two sets of data are evenly mixed indicates that the [\ion{N}{II}]/\Ha\ ratios are consistent between the two sets of observations. 

The median difference of the corrected and uncorrected \Ha\ emission is 
$ {\langle}\rm log \, (SFR_{H\alpha}/SFR_{H\alpha + [\ion{\rm N}{II}]}){\rangle} {=} {-0.14}$.
The scatter of the corrected and the observed \Ha~+~[\ion{\rm N}{II}] is $\delta (\rm log \, SFR_{H\alpha}/SFR_{H\alpha + [\ion{\rm N}{II}]}) = 0.13$. 
A linear-regression fit reveals that this correlation is a function of the measured $\rm SFR_{tot}$, with the difference being larger for larger SFRs (intercept $\alpha = -0.13 \pm 0.01$ and slope $\beta = -0.04 \pm 0.01$).
The negative slope is in agreement with the known positive correlation between the [\ion{N}{II}]/\Ha\ flux ratio and the SFR \citep[e.g.,][Figure \ref{fig:NII_Ha}]{2008ApJS..178..247K}.

\subsection{Extinction corrected \texorpdfstring{H$\alpha$}~~star-formation rates}
\label{sec:Ha_ext}

This section examines the effect of extinction on the \Ha-based SFR. 
We considered both the \Ha\, and (\Ha\ + [\ion{N}{II}]) luminosities; the latter in order to assess the effect of the systematic bias introduced by the [\ion{N}{II}] contamination in the \Ha\ photometry when reliable subtraction of this contribution is not possible. 
The SFRs were calculated as discussed in Section \ref{sec:Results}, and they are reported in the Appendix.
The extinction was calculated based on the Balmer decrement, the IRX, and the SED fits as discussed in Section \ref{sec:Extinction_indicators}.

\begin{figure*}
    \centering
    \includegraphics[width=\linewidth]{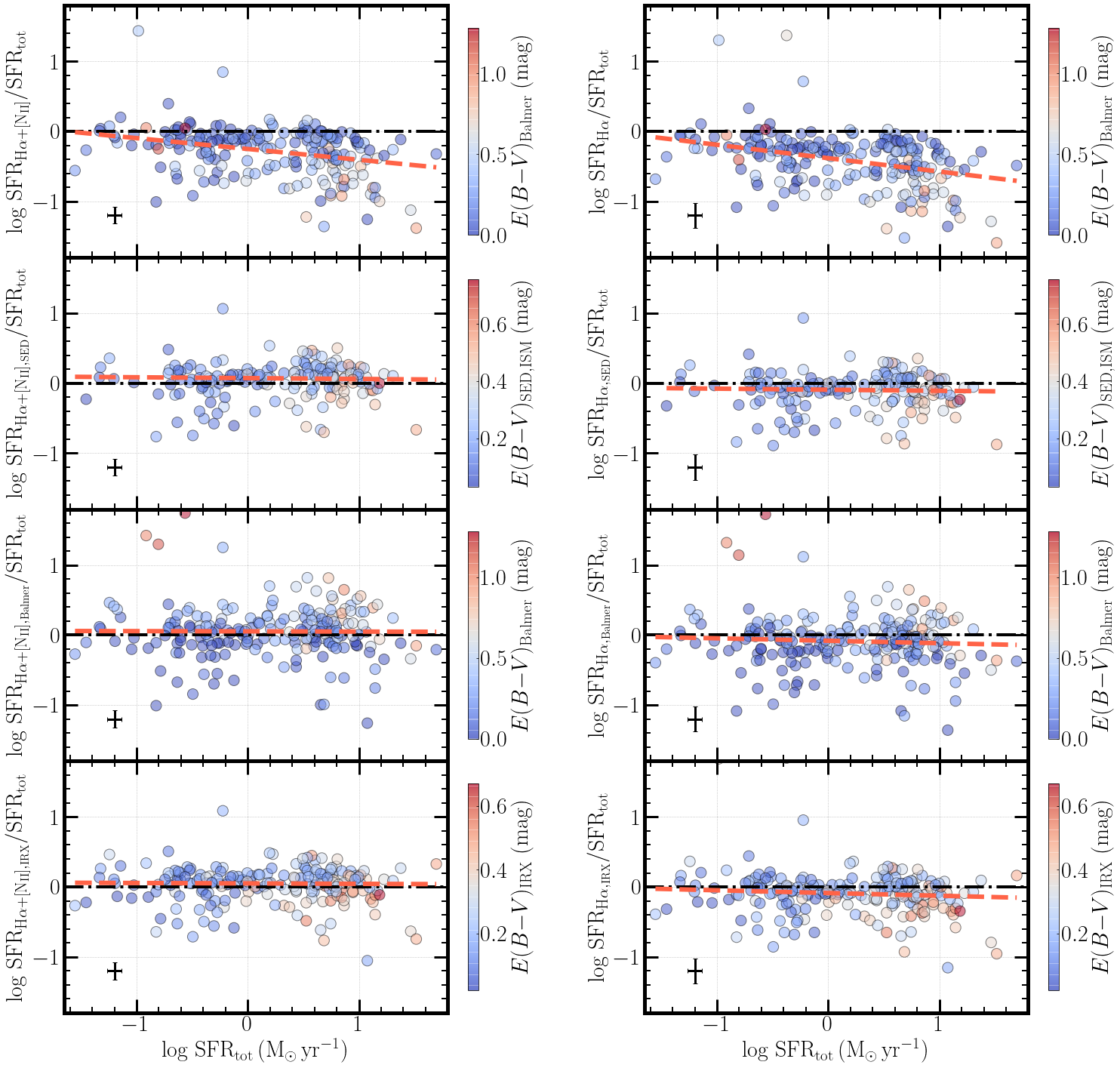}
    \caption{Comparisons between the ratio of SFRs derived by \Ha\ to $\rm SFR_{tot}$, not corrected for the [\ion{N}{II}] contribution (left column) and corrected (right column), as a function of $\rm SFR_{tot}$.
    From top to bottom, rows show SFRs: \textit{i}) not corrected for extinction,
    \textit{ii}) corrected by the $\rm SED_{ISM}$-based,
    \textit{iii}) corrected by the Balmer-based, and
    \textit{iv}) corrected by the IRX-based extinction.
    The black dashed-dotted and the red-dashed lines represent equality and the linear-regression fits, respectively (Table \ref{tab:Ha_to_tot_ext}).
    Each point represents a star-forming SFRS galaxy.
    Points are colour-coded based on the extinction used in each case respectively, indicated in the colour-bar at the right of each panel.
    The top panels show the Balmer decrement for each object for reference, even though no  extinction correction was applied.
    For clarity only extinction indicators which show reasonable agreement between \Ha\ and $\rm SFR_{tot}$ are shown.
    The black error bar at the bottom left of each panel indicates the median uncertainties.
    }
    \label{fig:Ha_ext_corr}
\end{figure*}

Figure \ref{fig:Ha_ext_corr} compares the uncorrected and extinction-corrected \Ha-based SFRs against $\rm SFR_{tot}$ for 201 star-forming SFRS galaxies.
The comparison with the sample with the SED-based extinction drops to 183 galaxies with SED-fits reduced $\chi^2 < 5$. 
The full array of comparisons is presented in Table \ref{tab:Ha_to_tot_ext}.
Table \ref{tab:Ha_to_tot_ext} reports the median values and standard deviation of the $\rm SFR_{H\alpha}/SFR_{tot}$ ratios, and the slopes and intercepts of the $\rm SFR_{H\alpha}/SFR_{tot}$ -- $\rm SFR_{tot}$ relations.  
The robust linear-regression fits are of the form:

\begin{eqnarray}
    \rm log \frac{SFR_x}{SFR_{tot}} = \alpha + \beta \, log SFR_{tot} \quad ,
    \label{eq:linear}
\end{eqnarray}
where x corresponds to the extinction indicators used to correct the \Ha\ flux.

\begin{table*}
    \centering
    \caption{Comparisons of SFR derived by \Ha\ with $\rm SFR_{tot}$.
    We considered \Ha\ SFRs corrected and not corrected for the [\ion{N}{II}] contribution and corrected or not for extinction using different extinction indicators.
    Fits are of the form of Eq. \ref{eq:linear}}
    \begin{tabular}{cccccc}
        Ext. correction & [\ion{N}{II}] correction & median & std. dev. & intercept & slope \\
         & & $\rm {\langle} log \frac{SFR_x}{SFR_{tot}} {\rangle} $ & $\rm \delta ( log \frac{SFR_x}{SFR_{tot}} ) $ & $\alpha$ & $\beta$\\
        \hline
        No & No &  $-0.20$ & $0.35$ & $-0.25 \pm 0.02$ & $-0.15 \pm 0.03$ \\
        $\rm SED_{ISM}$ & No & $0.08$ & $0.31$ & $0.07 \pm 0.02$ & $-0.01 \pm 0.02$\\        $\rm SED_{BC}$ & No & $0.29$ & $0.35$ & $0.26 \pm 0.02$ & $0.13 \pm 0.03$\\
        Balmer & No & $0.05$ & $0.42$ & $0.05 \pm 0.02$ & $0.00 \pm 0.03$\\
        IRX & No & $0.06$ & $0.30$ & $0.05 \pm 0.02$ & $-0.01 \pm 0.02$\\
        \hline
        No & Yes &  $-0.36$ & $0.37$ & $-0.39 \pm 0.02$ & $-0.19 \pm 0.03$ \\
        $\rm SED_{ISM}$ & Yes & $-0.07$ & $0.35$ & $-0.09 \pm 0.02$ & $-0.02 \pm 0.03$\\
        $\rm SED_{BC}$ & Yes & $0.14$ & $0.39$ & $0.10 \pm 0.02$ & $0.12 \pm 0.03$\\   
        Balmer & Yes & $-0.09$ & $0.43$ & $-0.08 \pm 0.02$ & $-0.03 \pm 0.03$\\
        IRX & Yes & $-0.08$ & $0.30$ & $-0.09 \pm 0.02$ & $-0.04 \pm 0.02$\\
    \end{tabular}
    \label{tab:Ha_to_tot_ext}
\end{table*}

As expected, \Ha\ and (\Ha\ + [\ion{N}{II}]), when not corrected for extinction, underestimate the SFR in all cases except for a few outliers.
When corrected for the [\ion{N}{II}] contribution, the \Ha-based SFRs show about $-0.15$ dex median difference with respect to the un-corrected.
This difference slightly increases for galaxies with high SFR, while galaxies with higher extinction tend to show even larger underestimation of the SFR.

The Balmer-decrement, IRX, and SED-ISM-based extinction all give similar corrections for \Ha-emission SFRs.
The comparison with $\rm SFR_{tot}$ results in a slightly sub-linear, almost flat slope considering the uncertainties, indicating that the correction is not strongly dependent on the SFR.
However, on average it slightly underestimates the total SFR by ${\sim} 0.1$ dex compared to $\rm SFR_{tot}$.
The correction based on the SED-fits BC component extinction leads to an overestimation of the SFRs that increases for higher-SFR galaxies, indicating this component overestimates the true extinction (Table \ref{tab:extinction}).
In the case of the combined BC~+~ISM $E(B{-}V)_{\rm SED}$, the comparison is strongly dependent on the SFR (slope $\beta = 0.3$), while the average SFR is overestimated by ${\sim} 0.4$ dex. 
For the rest of this work, we adopted the Balmer-decrement indicator for the extinction correction of the \Ha\ luminosities unless stated otherwise.

Based on the linear-regression fits with $\rm SFR_{tot}$ (Table \ref{tab:Ha_to_tot_ext}), one can infer $\rm SFR_{tot}$ through \Ha\ photometry when corrections for extinction and/or [\ion{N}{II}] contribution are not available by rewriting Equation \ref{eq:linear} as:
\begin{eqnarray}
    \rm SFR_{tot} = (SFR_x \, 10^{-\alpha})^{1/(\beta +1)} \quad .
    \label{eq:sfr_ha_to_tot}
\end{eqnarray}

\subsection{\texorpdfstring{8\,$\mu$m}\ ~emission as SFR indicator}
\label{sec:Discussion_SFR_8um}

\begin{figure}
    \centering
    \includegraphics[width=\columnwidth]{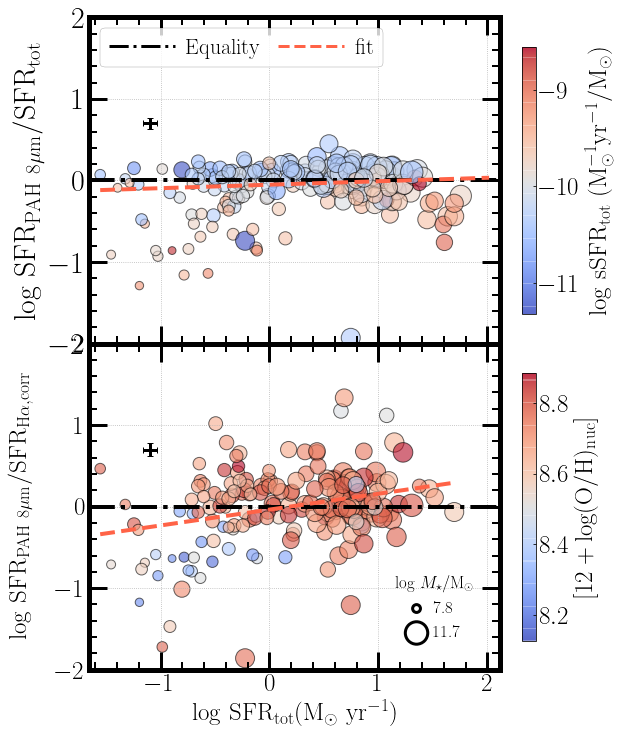}
    \caption{Top: $\rm SFR_{PAH~8~\mu m}$ to $\rm SFR_{tot}$ as a function of $\rm SFR_{tot}$.
    Galaxies are colour-coded based on the logarithm of their $\rm sSFR_{tot}$.
    Bottom: $\rm SFR_{PAH~8~\mu m}$ to $\rm SFR_{H\alpha}$ as a function of $\rm SFR_{tot}$.
    Galaxies are colour-coded based on their metallicity.
    In both panels the black dashed-dotted line represents the equality, and the red line the linear-regression fit (reported in Section \ref{sec:Discussion_SFR_8um}).
    Points size is a function of the galaxies' stellar mass.
    The black error bar at the top left of each panel indicates the median uncertainties.
    }
    \label{fig:PAH_tot}
\end{figure}

Figure \ref{fig:PAH_tot} shows the calibrations of SFRs derived through the 8\,$\mu$m PAH emission (Section \ref{sec:SFR}), $\rm SFR_{tot}$, and $\rm SFR_{H\alpha}$.
These comparisons involve 262 star-forming SFRS galaxies.
The comparison with $\rm SFR_{tot}$ shows very good agreement (linear-regression fit intercept $a = -0.05 \pm 0.02$ and slope $\beta = 0.04 \pm 0.02$).
However, the colour-coding of the points in Figure \ref{fig:PAH_tot} reveals that high sSFR galaxies systematically show lower $\rm SFR_{PAH,~8 \mu m}$ with respect to their $\rm SFR_{tot}$, as found by \cite{2019MNRAS.482..560M}. 
The comparison with \Ha-based SFRs (bottom panel of Figure \ref{fig:PAH_tot}) shows that the PAH~8~$\mu$m emission tends to underestimate the SFR (linear-regression intercept $a = -0.04 \pm 0.03$ and slope $\beta = 0.19 \pm 0.04$) in low-SFR and low-metallicity galaxies.

\subsection{SFRs based on WISE band-3 and band-4}
\label{sec:Discussion_SFR_WISE}

\begin{figure}
    \centering
    \includegraphics[width=\columnwidth]{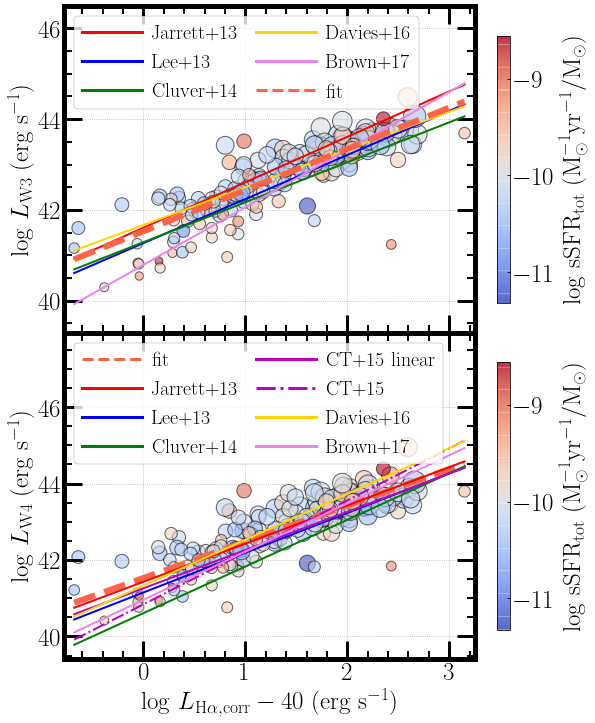}
    \caption{WISE band-3 (top) and band-4 (bottom) luminosities as a function of the Balmer-based extinction-corrected \Ha\ luminosity.
    Points are colour-coded and size-coded based on galaxies' sSFR and stellar mass respectively.
    Continuous lines represent calibrations from \protect\cite{2013AJ....145....6J,2013ApJ...774...62L,2014ApJ...782...90C,2015A&A...584A..87C,2016MNRAS.461..458D,2017ApJ...847..136B}.
    The error bars are smaller than the point sizes.
    The linear regression best-fit results are shown with a red dashed line and are reported in Table \ref{tab:LW_Ha}.}
    \label{fig:WISE_Lumin}
\end{figure}

Figure \ref{fig:WISE_Lumin} compares the WISE band-3 and band-4 luminosities 
derived from the WISE maps following the method similar to \cite{2019ApJS..245...25J}, as a function of the Balmer-decrement extinction-corrected \Ha\ luminosities of 262 star-forming SFRS galaxies.
The results of the linear-regression fits between ${\rm log} ~ L_{\rm WISE}$ and ${\rm log} ~L_{\rm H\alpha}$ are shown in Table \ref{tab:LW_Ha}.
These are in agreement with previous works although there is scatter ${\sim} 0.5$~dex for both bands.

\begin{figure}
    \centering
    \includegraphics[width=\columnwidth]{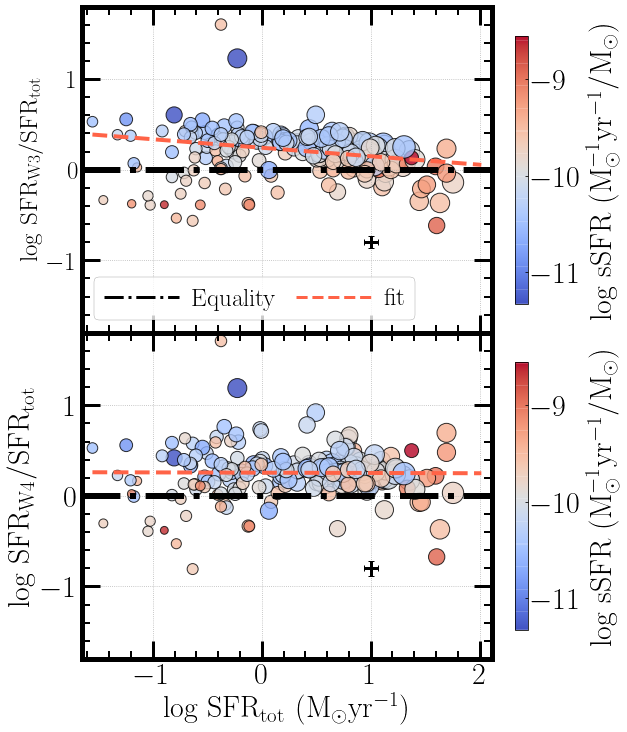}
    \caption{WISE band-3 and band-4 SFRs over $\rm SFR_{tot}$ as a function of $\rm SFR_{tot}$.
    Points are colour-coded and size-coded based on galaxies' sSFR and stellar mass respectively.
    The red dashed lines represents the linear-regression fits (Table \ref{tab:LW_Ha}), and the black dashed-dotted lines represents equality.
    The black error bar at the bottom right of each panel indicates the median uncertainties.}
    \label{fig:WISE_SFR}
\end{figure}

Figure \ref{fig:WISE_SFR} shows the comparison between WISE band-3 and band-4 SFRs, based on the calibration from \cite{2017ApJ...850...68C}, with $\rm SFR_{tot}$.
This calibration results in increased SFR by ${\sim} 0.35$~dex for both WISE bands with respect to $\rm SFR_{tot}$.

\begin{table}
    \centering
    \caption{Results of the linear-regression fits between ${\rm log} ~ L_{\rm WISE}$ as a function of ${\rm log} ~L_{\rm H\alpha} - 40$ and $\rm log ~ SFR_{\rm WISE}/SFR_{tot}$ as a function of $\rm log ~SFR_{tot}$.}
    \begin{tabular}{ccc}
         & intercept & slope \\
         & $\alpha$ & $\beta$\\
        \hline
        $L_{\rm W3}$--$L_{\rm H\alpha}$ & $41.52 \pm 0.06$ &  $0.92 \pm 0.04$\\
        $L_{\rm W4}$--$L_{\rm H\alpha}$ & $41.52 \pm 0.06$ &  $0.96 \pm 0.04$\\
        \hline
        $\rm (SFR_{W3}/SFR_{tot})$--$\rm SFR_{tot}$ & $0.24 \pm 0.01$ &  $-0.09 \pm 0.02$\\
        $\rm (SFR_{W4}/SFR_{tot})$--$\rm SFR_{tot}$ & $0.26 \pm 0.01$ &  $-0.01 \pm 0.02$\\
    \end{tabular}
    \label{tab:LW_Ha}    
\end{table}

\subsection{Combinations of \texorpdfstring{H$\alpha$}~, 24, and PAH \texorpdfstring{8~$\mu$m}~ emission as hybrid star-formation indicators}
\label{sec:SFR_Hybrid}

\Ha\ traces the Lyman-continuum UV photons produced by stellar populations younger than those traced by the typical UV bands that can be probed directly \citep[e.g.,][]{2012AJ....144....3L,2012ARA&A..50..531K,2014A&A...571A..72B,2020MNRAS.494.5967K,2020MNRAS.498..235H}.
The latter include emission of B stars.
However, the main limitation of \Ha\ (and in general all optical and UV) emission is that it is affected by extinction, which often is hard to estimate reliably.

Hybrid indicators like the combination of 24~$\mu$m and 8~$\mu$m with \Ha\ or [\ion{O}{II}] emission \citep[e.g.,][]{2007ApJ...666..870C,2009ApJ...703.1672K} account for both the dust-absorbed and unabsorbed radiation.
The 24~$\mu$m emission is unaffected by extinction and traces the reprocessed emission of young stellar populations \citep[ages $ \lesssim 200$~Myr; e.g.,][]{2012ARA&A..50..531K,2014A&A...571A..72B, 2016A&A...589A.108C,2020MNRAS.494.5967K}.
In order to estimate the SFR from the above combinations we adopted the conversions of  \cite{2009ApJ...703.1672K}:

\begin{eqnarray}
    \rm SFR = 7.9 \, \frac{\textit{L}_{H\alpha} + 0.02 \, \textit{L}_{\rm 24\mu m}}{10^{42} ~ \erg} \quad ,
\end{eqnarray}
where $L_{24\mu m}$ refers to $\nu L_{\nu}$ at 24~$\mu$m, and 
\begin{eqnarray}
    \rm SFR = 7.9 \, \frac{\textit{L}_{H\alpha} + 0.011 \, \textit{L}_{\rm PAH~8\mu m}}{10^{42} ~ \erg} \quad ,
\end{eqnarray}
where $L_{\rm PAH~8\mu m}$ refers to $\nu L_{\nu}$ at 8~$\mu$m (Eq. \ref{eq:Helou}) from PAHs.

\begin{figure}
    \centering
    \includegraphics[width=\columnwidth]{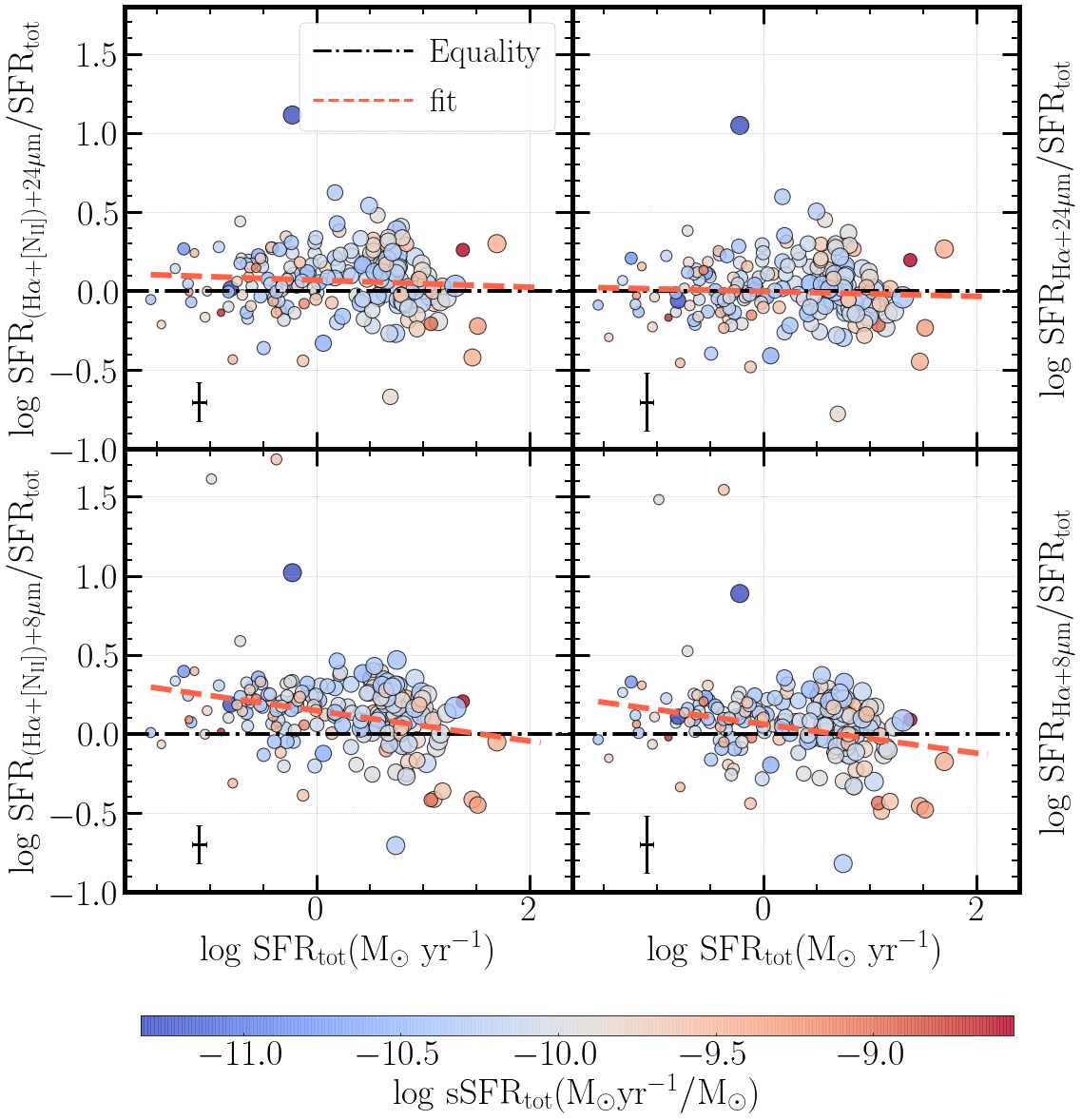}
    \caption{\Ha\ + 24~$\mu$m (top panels) and \Ha\ + 8~$\mu$m (bottom panels) SFRs as a function of $\rm SFR_{tot}$ with (right panels) and without (left panels) correction for the contribution of the [\ion{N}{II}] emission.
    The points are colour-coded and size-coded based on galaxies' sSFR and stellar mass respectively.
    The black dashed and red dashed-dotted lines represent equality and the linear-regression fits (Table \ref{tab:24_Ha}), respectively.
    The black error bar at the bottom left of each panel indicates the median uncertainties.}
    \label{fig:SFR_hybrids}
\end{figure}

Figure \ref{fig:SFR_hybrids} compares the 24~$\mu$m~+~\Ha\ and the PAH~8~$\mu$m~+~\Ha\ SFR indicators to $\rm SFR_{tot}$ while also investigating the effect of the [\ion{N}{II}] contribution.
This comparison involves 247 star-forming SFRS galaxies which have 24~$\mu$m, \Ha, and optical spectral observations. 
The linear-regression fit results are given in Table \ref{tab:24_Ha}.

The 24~$\mu$m~+~\Ha\ SFRs show in both cases (with or without the [\ion{N}{II}] correction) only small offsets compared to $\rm SFR_{tot}$ and no evidence for correlation with SFR (slope of the linear-regression fit $\beta \eqsim 0$).
The absolute differences show that the 24~$\mu$m~+~(\Ha\ + [\ion{N}{II}]) slightly overestimates (by $0.07$~dex) the SFR, while the 24~$\mu$m~+~\Ha\ is in excellent agreement with $\rm SFR_{tot}$.
The scatter is slightly higher in the case of the corrected for the [\ion{N}{II}] contribution $\rm SFR_{24\mu m~+~H\alpha}$.

The PAH~8~$\mu$m~+~\Ha\ SFRs show worse agreement in comparison to the 24~$\mu$m as in both cases the slopes of the linear-regression fits are negative. 
This is driven by low SFR galaxies, which are systematically above the equality line, and galaxies with SFR~$\simeq$10--50~\msunpyr, which show increased sSFR but a systematic deficit in their \Ha~+~$\rm PAH_{8~\mu m}$ SFR.

\begin{table*}
    \centering
    \caption{Results of the comparisons of the SFRs derived by \Ha\ corrected and not-corrected for the [\ion{N}{II}] contribution in combination with 24~$\mu$m and PAH emission and by the SED fits, all compared to $\rm SFR_{tot}$.}
    \begin{tabular}{ccccc}
         & median & std. & intercept & slope \\
         & $\rm {\langle} log \frac{SFR_x}{SFR_{tot}} {\rangle} $ & $\rm \delta ( log \frac{SFR_x}{SFR_{tot}} ) $ & $\alpha$ & $\beta$\\
        \hline
        24~$\mu$m + (\Ha\ + [\ion{N}{II}]) & $0.06$ &  $0.28$ & $0.07 \pm 0.01$ & $-0.02 \pm 0.02$\\
        24~$\mu$m~+~\Ha\ & $-0.02$ & $0.27$ & $0.0 \pm 0.01$ & $-0.02 \pm 0.02$\\
        $\rm PAH_{8~ \mu m}$+(\Ha\ + [\ion{N}{II}]) & $0.15$ & $0.28$ & $0.15 \pm 0.01$ & $ -0.10 \pm 0.02$\\
        $\rm PAH_{8~ \mu m}$~+~\Ha\ & $0.07$ &  $0.25$ & $0.06 \pm 0.01$ & $-0.09 \pm 0.02$\\
        SED & $0.02$ &  $0.42$ & $-0.02 \pm 0.01$ & $0.04 \pm 0.02$\\
    \end{tabular}
    \label{tab:24_Ha}
\end{table*}

\section{Discussion}
\label{sec:Discussion}

\subsection{Extinction-corrected \texorpdfstring{H$\alpha$}~ emission and the contribution of the \texorpdfstring{[\ion{N}{II}]}~ emission.}
\label{sec:Discussion_Ha_NII}

Through the evolution of the stellar populations, the ISM is enriched with metals, which can form complex molecules and dust under appropriate conditions.
In fact, the attenuation laws describing the effect of extinction as a function of wavelength are a complex function that depends on the properties of the dust grains as well as the spatial distribution of dust in the ISM with respect to the stars \citep[e.g.,][]{1994ApJ...429..582C,2000ApJ...533..682C}.
The attenuation laws can vary significantly in different galaxies \citep[e.g.,][]{2018A&A...619A.135B,2018ApJ...859...11S}. 

Because star formation requires the presence of gas (and dust), the recently born stars are often embedded in regions with large dust and gas column density.
Thus, their emission can be partially or completely absorbed.
The total absorption is a combination of the absorption at the sites of star formation and the intervening dust in the ISM along the line-of-sight  \citep[e.g.,][]{2000ApJ...539..718C,2011MNRAS.417.1760W,2014ApJ...788...86P,2015ApJ...806..259R}.
SFR is, directly or indirectly, measured through these young stars' emission.
Therefore, in order to infer the correct SFR, one must account for the extinction. 

The Balmer-decrement, IRX, and SED-ISM-based extinctions are on average in good agreement (Table \ref{tab:extinction}; Figure \ref{fig:corner_ext}).
The mode of the ratio of the color excess between the two methods $R_{\rm EBV}= {\langle}E(B{-}V)_{\rm IRX}/E(B{-}V)_{\rm Balmer}{\rangle} = 0.90^{+0.75}_{-0.37}$ for 211 star-forming SFRS galaxies with $3\sigma$ detections for both extinction indicators. 
\cite{2019ApJ...886...28Q} found $\rm \textit{R}_{EBV}=0.51$ based on a sample of SDSS-DR10 galaxies \citep[][]{2014ApJS..211...17A}.
However, if we consider \Ha\ measurements from the nuclear regions of the SFRS galaxies (but integrated IR and UV photometry for the IRX), $R_{\rm EBV}=0.54^{+0.25}_{-0.17}$ in agreement with \cite{2019ApJ...886...28Q}.
Therefore, the difference may be due to the fact that \cite{2019ApJ...886...28Q} Balmer-decrement extinctions were derived from nuclear regions of galaxies (as for all SDSS spectra) instead of their integrated average.
However, as indicated from the 68\% confidence intervals on $R_{\rm EBV}$, both nuclear and wide-region comparisons show considerable scatter.

The comparison of the \Ha-derived SFRs with $\rm SFR_{tot}$  (Figure \ref{fig:Ha_ext_corr}) offers insight on the extinction indicators.
The [\ion{N}{II}]-corrected \Ha\ SFRs are in excellent agreement with $\rm SFR_{tot}$ when corrected with Balmer, IRX, and SED-ISM-based extinction considering the uncertainties.
The fact that $\rm SFR_{tot}$ is on average slightly higher compared to the \Ha-based SFRs can be attributed to the fact that the FUV and FIR also trace older stellar populations (up to $200$~Myr) compared to \Ha\ emission (${\leq} 10$~Myr).
The fact that the difference with $\rm SFR_{tot}$ is slightly increased for high-SFR galaxies may be driven by the inability of the extinction indicators to trace extinction in galaxies with the highest attenuation.
While for the bulk of the sample there is good agreement between the extinction-corrected $\rm SFR_{H\alpha}$ and $\rm SFR_{tot}$, there is a tail towards lower values of the $\rm SFR_{H\alpha}/SFR_{tot}$ distribution which could be the result of spatial variations of the extinction on subgalactic scales leading to leakage of UV photons.
These photons are accounted for in $\rm SFR_{tot}$, but they do not contribute to $\rm SFR_{H\alpha}$.

\Ha\ imaging is one of the easiest ways to measure and map the SFR of a local-Universe galaxy and certainly the closest to the instantaneous SFR \citep[e.g.,][]{2012ARA&A..50..531K,2012AJ....144....3L}.
However, as the cost of spectroscopic observations is usually higher compared to imaging, it is common to lack measurements of the Balmer decrement or the [\ion{N}{II}]/\Ha\ ratio.
\Hb\ imaging observations require around nine times longer exposure time compared to \Ha\ in order to obtain observations of similar signal-to-noise ratio.
Also, there is not always available photometric coverage in FIR and FUV bands, which are required for the IRX extinction and for SED fits.
Therefore, it is useful to calibrate the SFR inferred from the \Ha\ luminosity uncorrected for extinction or the [\ion{N}{II}]-lines contribution.

Table \ref{tab:Ha_to_tot_ext} gives the correlations between $\rm SFR_{tot}$ and the SFR from the \Ha\ luminosity uncorrected for extinction (with and without correction for the [\ion{N}{II}] contribution).
These correlations together with Equation \ref{eq:sfr_ha_to_tot} can be used to infer the intrinsic total SFR.
Table \ref{tab:Ha_to_tot_ext} also gives similar correlations between $\rm SFR_{tot}$ and \Ha-based SFR when different extinction-correction methods are applied.
These together with Equation \ref{eq:sfr_ha_to_tot} can be used to remove any  biases introduced by individual extinction corrections (although most are small as seen from the  results reported in Table \ref{tab:Ha_to_tot_ext}).

\begin{figure}
    \centering
    \includegraphics[width=\columnwidth]{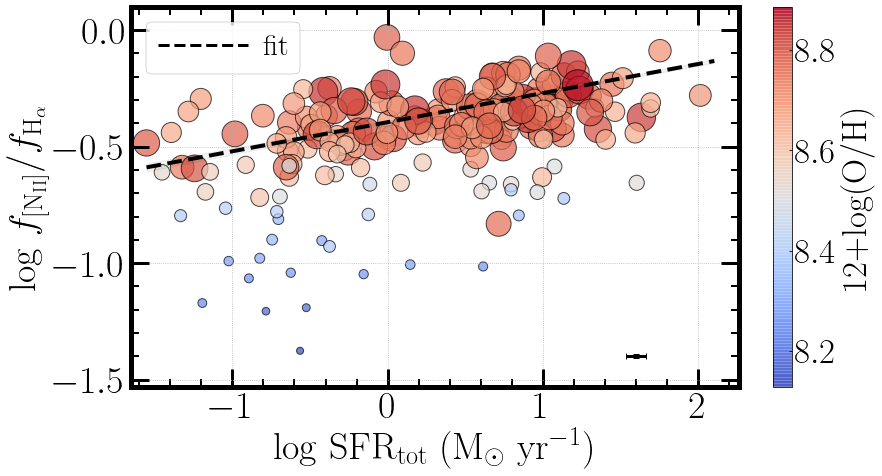}
    \caption{
    SFRS star-forming galaxies [\ion{N}{II}]/\Ha\ flux ratio as a function of $\rm SFR_{tot}$, in logarithmic space.
    Points are colour-coded based on their metallicity, and size-coded based on their stellar mass.
    The black-dashed line represents the linear-regression fit.
    The black error bar at the bottom right indicates the median uncertainties.
    The linear regression best-fit result is reported in Section \ref{sec:Discussion_Ha_NII}.}
    \label{fig:NII_Ha}
\end{figure}

The \Ha\ + [\ion{N}{II}] extinction-corrected SFR is also in agreement with $\rm SFR_{tot}$ considering the uncertainties.
This agreement holds for the full range of SFRs.
However \Ha-based SFRs not corrected for the [\ion{N}{II}] contribution show on average about $0.14$~dex higher values compared to the corrected ones (Section \ref{sec:Ha_ext}).
Figure \ref{fig:NII_Ha} shows a positive correlation between the [\ion{N}{II}]/\Ha\ ratio with SFR. 
A fit of ${\rm log} ~ f_{[\ion{N}{II}]}/f_{\rm H\alpha}$ against $\rm log~ SFR_{tot}$ gives a slope of $\beta = 0.12 \pm 0.01$ and  intercept $\alpha = -0.44 \pm 0.01$. 
The positive correlation is because galaxies with higher SFRs tend to have higher metallicity and a larger percentage of very-young stellar populations. 
Both factors result in increased excitation of the gas \citep[e.g.,][]{2008ApJ...681.1183K}. 
Galaxies deviating from the general relation are mainly dwarf galaxies with low metallicities (see also Section \ref{sec:Discussion_Metal}).
The scatter in high-metallicity galaxies can be attributed to the varying ionization degree of the gas, which depends on the local star-formation conditions (e.g., BPT diagrams).
The [\ion{N}{II}]/\Ha\ ratio can vary up to 0.5~dex for ionization parameter\footnote{The ionization parameter represents the intensity of the ionizing field with respect to the gas density.} $q$ varying from $10^7$--$\rm 10^8 \, cm~s^{-1}$ and constant metallicity \citep[although metallicity also plays a role; see Figure 9 of ][]{2008ApJ...681.1183K}.
Nonetheless, the relatively small contribution of the [\ion{N}{II}] line with respect to \Ha\ emission results in a weak dependence of the total SFR as shown in the slopes of the comparisons (Table \ref{tab:Ha_to_tot_ext}).

\subsection{PAHs emission as SFR indicator}
\label{sec:Discussion_PAH}

Both 8 and 12 $\mu$m emission trace PAH molecules.
Figures \ref{fig:PAH_tot} and \ref{fig:WISE_SFR} show that galaxies with high sSFR (a proxy for the star-formation intensity and the stellar population age) have a deficit in PAH emission.
This behavior with respect to $\rm SFR_{tot}$ and the galaxies' sSFR is similar for both 8 and 12 $\mu$m emission. 
This is evident in both ends of high or low SFRs and stellar masses. 
As discussed by \cite{2019MNRAS.482..560M}, this trend can be attributed to a) the fact that the PAHs are excited in photo-dissociation regions which are in the surface of the star-forming bubbles, while the \Ha\ emission comes from their volume, and b) the destruction of PAH molecules caused by intense UV radiation field in highly star-forming galaxies \citep[e.g.,][]{2014A&A...566A.136M}.
The latter is supported by the fact that the deficit of PAHs with respect to the TIR emission is higher in galaxies where star formation takes place in more compact environments \citep[e.g.,][]{2011A&A...533A.119E,2011ApJ...741...32D}.

Figure \ref{fig:PAH_tot} shows that a deficit in the PAH emission in the low-SFR regime is associated with low-metallicity galaxies. 
In this case the deficit is due to the lower dust and PAH molecule abundance in dwarf star-forming galaxies which are characterized by low metallicities.
In contrast, $\rm SFR_{tot}$ is not affected by dust deficiency in the low-metallicity environment because $\rm SFR_{tot}$ accounts for the unobscured star-forming activity.

\subsection{SFRs from SED fits}
\label{sec:Discussion_SFR_SEDs}

\begin{figure}
    \centering
    \includegraphics[width=\columnwidth]{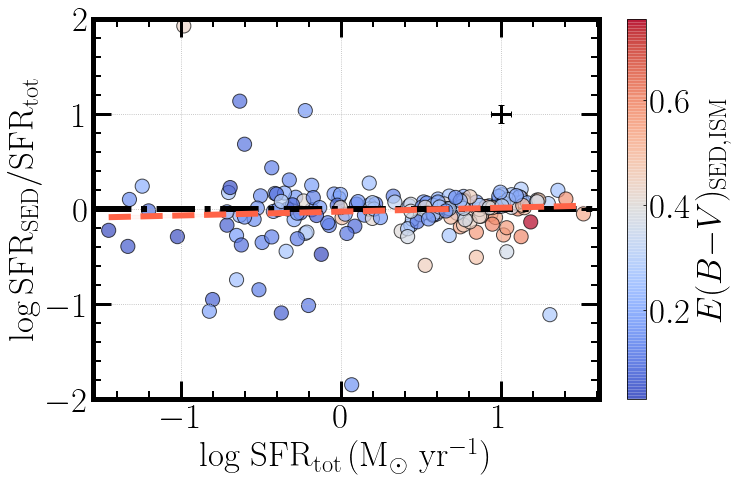}
    \caption{
    The logarithm of the $\rm SFR_{SED}/SFR_{tot}$ ratio as a function of $\rm SFR_{tot}$.
    This comparison includes 183 star-forming SFRS galaxies with good quality fits.
    The points are colour-coded according to the SED-ISM component extinction.
    The red dashed and black dashed-dotted lines represent their linear-regression fit (Table \ref{tab:24_Ha}) and equality respectively.
    The black error bar at the top right indicates the median uncertainties.}
    \label{fig:SED_tot}
\end{figure}

The SED-derived SFRs are considered to be close to the true SFRs because they model the galaxies' emission using photometric information in a wide range of wavelengths, explicitly accounting for the effects of extinction.
Figure \ref{fig:SED_tot} compares the ratio of $\rm SFR_{SED}$ and $\rm SFR_{tot}$ as a function of $\rm SFR_{tot}$.
It shows an overall good agreement, which is excellent for SFR~${>}~1$~\msunpyr.
In this range of  SFRs the scatter is minimized.
However, in the regime of low SFRs the comparison shows significantly increased dispersion.
This can be attributed to spatial variations in the age of the stellar populations \citep[e.g.,][]{2020MNRAS.498..235H,2020MNRAS.494.5967K}, stochasticity in the extremely small SFRs \citep[e.g.,][]{2012ARA&A..50..531K}, and/or extinction variations between the different star-forming regions within a galaxy \citep[e.g.,][]{2020ARA&A..58..529S}.
As a result, different regions may dominate the emission in different wavelengths, an effect that cannot be  taken into account effectively in an SED analysis framework \citep[although some efforts in this direction have been made; e.g.,][]{2008MNRAS.388.1595D}.
This effect becomes more important for low SFR, the IR luminosity of which may be dominated by a few individual star-forming regions, while their visible, NIR, and UV emission may arise from other regions hosting older and less obscured populations.

\subsection{Hybrid SFR indicators}
\label{sec:Discussion_SFR_Hybrids}

The hybrid 24~$\mu$m~+~\Ha\ SFRs show excellent agreement with the FIR~+~FUV emission based SFR (Figure \ref{fig:SFR_hybrids}; \citealt{2009ApJ...703.1672K}).
This agreement is not significantly affected by the [\ion{N}{II}] contribution, due to a combination of two effects:
a) the [\ion{N}{II}]/\Ha\ ratio is correlated with SFR (Figure \ref{fig:NII_Ha}) with the [\ion{N}{II}] contribution being relatively lower for low SFRs, and
b) the \Ha/24~$\mu$m flux ratio decreases for increasing SFRs.
These effects cancel out, resulting to a close to flat slope in both the corrected and uncorrected fits.
However, the average [\ion{N}{II}] contribution increases the inferred SFR by ${\sim} 0.1$~dex (Table \ref{tab:24_Ha}). 
Overall, the hybrid 24~$\mu$m~+~\Ha\ SFR is an excellent alternative to the FIR~+~FUV, even when
it is not possible to correct for the contribution of the [\ion{N}{II}] emission.

The hybrid 8~$\mu$m~+~\Ha\ SFRs show decent agreement with $\rm SFR_{tot}$ (Figure \ref{fig:SFR_hybrids}), but this indicator tends to underestimate the SFR in the high-SFR regime.
As discussed in Section \ref{sec:Discussion_SFR_8um}, this is caused by the fact that the PAH 8~$\mu$m emission shows a deficit in intensively star-forming galaxies, failing to account for the reduction of the \Ha\ flux due to absorption.
Similarly to the 24~$\mu$m~+~\Ha\ SFR indicator, no [\ion{N}{II}] correction causes an average increase of ${\sim} 0.1$ dex in the SFR measurement.

\subsection{Radio emission as SFR indicator}
\label{sec:Discussion_SFR_radio}

The 1.4\,GHz radio continuum emission traces synchrotron emission produced by the interaction of relativistic electrons and cosmic rays produced in supernovae remnants with the galactic magnetic field. 
The lifetime of relativistic electrons depends on their two main energy-loss mechanisms: synchrotron emission and inverse-Compton scattering of photons in the radiation field of the galaxy.
Assuming a range of galactic magnetic field from $B = $ 1 up to 25 $\mu$Gauss \citep[e.g.,][]{2010ASPC..438..197F}, synchrotron cooling timescales are 5--160~Myr (Equation 18 from \citealt{2010ApJ...717....1L}).
The inverse-Compton cooling timescale depends on the magnetic field and the radiation-field density, which can be parametrized by the SFR surface density $\rm \sum_{SFR}$ (Equation 23 off \citealt{2010ApJ...717....1L}).
Based on the SFRS galaxies $\rm \sum_{SFR}$ distribution and assuming $B=10~\mu$Gauss, the inverse-Compton timescales can be 4~kyr--15~Myr. 
The inverse-Compton cooling dominates the energy losses of relativistic electrons in highly star-forming galaxies.
Therefore, the radio emission traces timescales similar to or shorter than the lifetimes of massive stars and probes similar stellar populations as the \Ha\ emission.
In addition, it has the benefit that is not affected by extinction, and it gives a complementary view of star-formation because it probes different processes than those producing the IR and 24~$\mu$m (heated dust) or the \Ha\ emission (gas ionized by UV).

\begin{figure*}
    \centering
    \includegraphics[width=\textwidth]{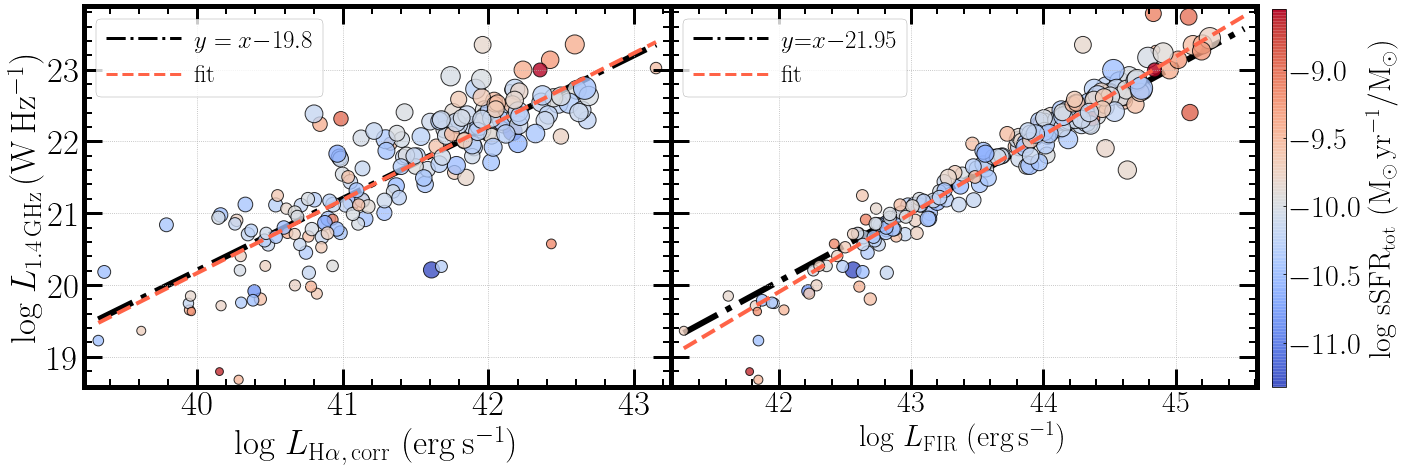}
    \caption{
    The logarithm of the $L_{\rm 1.4~GHz} ~ ({\rm W~Hz^{-1}})$ as a function of the [\ion{N}{II}] and (Balmer-decrement) extinction-corrected $L_{\rm H\alpha}$ (left panel), and the $L_{\rm FIR}$ (right panel).
    This comparison involves 233 star-forming SFRS galaxies with radio and ${\rm H\alpha}$ photometry.
    Points are colour-coded depending on their $\rm sSFR_{tot}$ and size-coded based on their stellar mass. 
    The red dashed line represent the linear-regression fit (reported in Section \ref{sec:Discussion_SFR_radio}), and the black dashed-dotted line shows a linear correlation for reference.
    The error bars are smaller than the point sizes.}
    \label{fig:Lumin_Radio_Ha}
\end{figure*}

Figure \ref{fig:Lumin_Radio_Ha} compares the 1.4\,GHz luminosity \citep[adopted from][]{2011PASP..123.1011A} with the \Ha\ and the FIR luminosities.
Many studies have shown a tight correlation between radio and IR luminosities \citep[e.g.,][]{1992ARA&A..30..575C,1999MNRAS.302..632B,1999ApJ...517..148F,2003ApJ...586..794B}.
Unsurprisingly, a tight correlation is also found for the SFRS sample, where a linear-regression fit between the 1.4\,GHz and FIR luminosities shows excellent agreement: ${\rm log}~L_{\rm 1.4\,GHz} ~ ({\rm 10^{22}~ W~Hz^{-1}}) =
(-3.99 \pm 0.58) + (1.09 \pm 0.01) ~ {\rm log} ~ L_{\rm FIR} ~ ({\rm erg~s^{-1}})$ with scatter $\delta ({\rm log } \frac{L_{\rm 1.4\,GHz}}{L_{\rm FIR}}) = 0.26$.
The comparison between $L_{\rm 1.4\,GHz}$ and $L_{\rm H\alpha}$ also shows excellent agreement [${\rm log}~L_{\rm 1.4\,GHz} ~ ({\rm 10^{22}~ W~Hz^{-1}}) = (1.09 \pm 1.7) + (1.03 \pm 0.04) ~ {\rm log} ~ L_{\rm H\alpha} ~ ({\rm erg~s^{-1}})$] but with large scatter $\delta ({\rm log }\frac{L_{\rm 1.4\,GHz}}{L_{\rm H\alpha}}) = 0.52$.

As discussed by \cite{2003ApJ...586..794B}, the linearity and tightness in the FIR/radio correlation can be considered as a conspiracy.
The FIR emission fails to trace star formation in low-luminosity dust-deficient galaxies, while in the more massive (and generally higher metallicity) galaxies, FIR emission can be augmented by contribution from older stellar populations. 
The radio emission also underestimates star formation in faint galaxies due to decreased non-thermal radio emission efficiency in these objects.
This has been attributed to either cosmic-ray escape losses at low SFRs \citep[e.g.,][]{1990JPhG...16.1409C,2003ApJ...586..794B,2010ApJ...717....1L} or possibly stronger magnetic fields in higher SFR galaxies \citep[e.g.,][]{2017ApJ...836..185T}.
\cite{2017MNRAS.466.2312D} report non-linear relation of the form ${\rm SFR} \propto L_{\rm 1.4~GHz}^\gamma$ with $\gamma  = 0.75$.
However, other studies have found a plethora of values for $\gamma$, ranging e.g., between 0.77 and 1.06
\citep[][]{1992ApJ...401...81P}
depending on the sample and the reference SFR indicator.
Based on Eq. \ref{eq:SFR-Ha} and adopting a linear relation between the radio
1.4\,GHz and \Ha\ luminosities, we propose a conversion from 1.4~GHz luminosity to SFR.
\begin{eqnarray}
    \frac{\rm SFR_{1.4~GHz}}{(\msunpyr)} = 3.38 \, \frac{{L}_{\rm 1.4~GHz}}{({\rm 10^{22} ~ W~{Hz}}^{-1})} \quad .
	\label{eq:SFR-1.4GHz}
\end{eqnarray}

\begin{figure}
    \centering
    \includegraphics[width=\columnwidth]{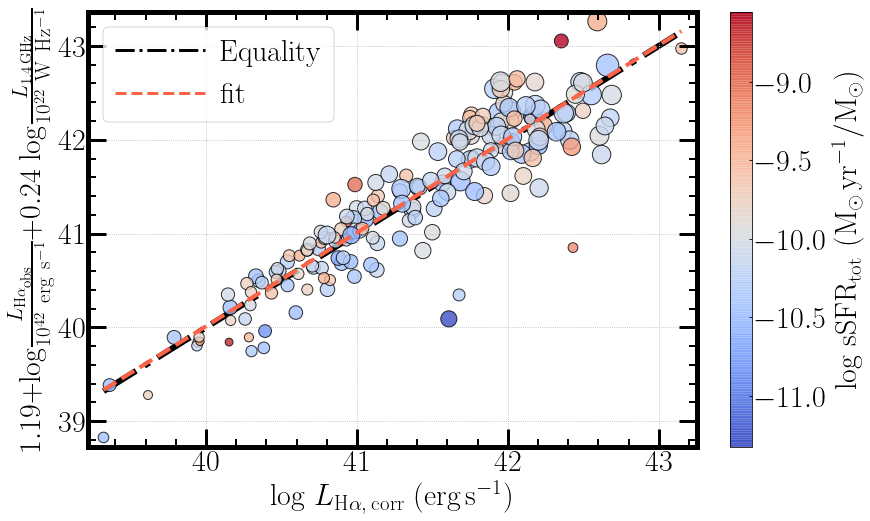}
    \caption{Combined observed \Ha\ (corrected for the [\ion{N}{II}] contribution but not for extinction) and radio 1.4~GHz luminosity as a function of the [\ion{N}{II}] and (Balmer-decrement) extinction-corrected \Ha\ luminosity.
    Points are colour-coded based on their $\rm sSFR_{tot}$, and size-coded based on their stellar-mass.
    The black dashed-doted line and the red dashed line represent the equality and the linear-regression fit respectively.
    The error bars are smaller than the point sizes.}
    \label{fig:Lumin_Radio_Ha_comb}
\end{figure}

Following \cite{2009ApJ...703.1672K}, we calibrated the composite SFR indicator combining the radio 1.4~GHz and the {\em not} corrected for extinction \Ha\ luminosities in analogy to the 24~$\mu$m and 8~$\mu$m~+~\Ha\ hybrid SFR indicators.
Combining with Eq. \ref{eq:SFR-Ha}: 
\begin{eqnarray}
    \begin{split}
    {\rm log}~\frac{\rm SFR}{(\msunpyr)} = 1.19 +
      {\rm log} ~  \frac{{L}_{\rm H\alpha, obs}}{({\rm 10^{42} ~ erg~s}^{-1})}\\ 
      + ~ 0.24 \, {\rm log} ~ \frac{{L}_{\rm 1.4~GHz}}{({\rm 10^{22} ~ W~{Hz}^{-1}})} \quad .
    \end{split}
\end{eqnarray} 
Figure \ref{fig:Lumin_Radio_Ha_comb} shows the comparison between the combined \Ha~+~1.4~GHz and the extinction-corrected $L_{\rm H\alpha}$ luminosities.

The slope in the SFR--$L_{\rm 1.4~GHz}$ relation depends on the SFR indicator used for the calibration.
The tight correlation between the FIR and 1.4~GHz luminosities results in a linear slope, but the FIR emission is not accurately tracing star-formation in low-luminosity, dust-deficient galaxies.
The combination of UV~+~IR emission as SFR indicator ($\rm SFR_{tot}$) results in a non-linear calibration because the UV used to correct for the missing IR emission in the low-SFR galaxies also traces stellar populations with older ages (up to 200~Myr) compared to the radio emission.
Instead, \Ha\ gives a slope close to linear because it traces the same stellar populations ages as the radio.
The minimum sSFR for the SFRS galaxies ($\rm log ~ sSFR_{H\alpha}  = -11.44 ~ M_\odot ~ yr^{-1} / M_\odot$ in NGC~4491) is higher than $\rm log ~ sSFR \gtrsim -12$, the limiting sSFR below which older stellar populations contribute to the UV ionization field \citep[][]{2020MNRAS.494.5967K}. 
This ensures that the \Ha-based SFRs trace indeed the youngest stellar populations, which are expected to correlate with the radio emission.

\begin{figure}
    \centering
    \includegraphics[width=\columnwidth]{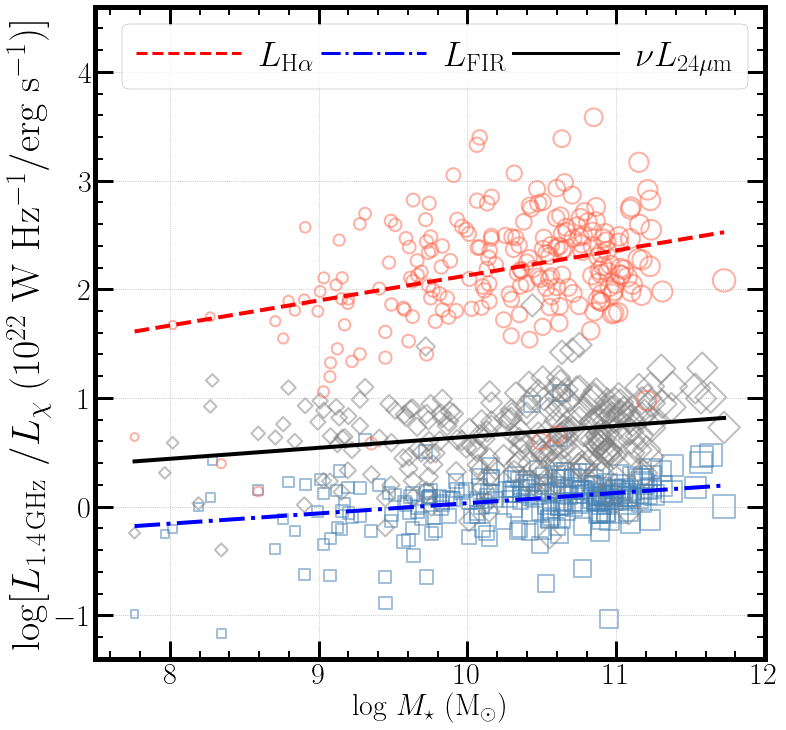}
    \caption{The ratio between the radio 1.4~GHz, and the \Ha\ (red circles), FIR (black diamonds; calculated following \protect\citealt{1988ApJS...68..151H}), and 24$\mu$m (blue squares) luminosities as a function of the galaxies' stellar mass.
    The points size is also a function of the galaxies' stellar mass.
    The dashed red, black, and dashed-dotted blue lines, represent their respective linear-regression fits (reported in Section \ref{sec:Discussion_SFR_radio}).
    The error bars are smaller than the point sizes.}
    \label{fig:Lumin_Radio_Ha_mass}
\end{figure}

In order to investigate the increased dispersion in the radio/\Ha\ relation, Figure \ref{fig:Lumin_Radio_Ha_mass} compares the radio-to-\Ha\ ($q_{\rm H\alpha}$), radio-to-FIR ($q_{\rm IR}$), and radio-to-24~$\mu$m ($q_{\rm 24 \mu m}$) luminosity ratios as a function of the galaxies' stellar mass.
Low-mass galaxies show on average a deficit of radio to \Ha\ emission compared to higher-mass galaxies.
This can be attributed to two mechanisms: 
a) the weak gravitational field in lower-mass galaxies makes them more prone to the formation of galactic-scale winds and higher escape losses of relativistic electrons, and
b) dwarf galaxies do not have well-formed spiral arms and disk resulting in weaker large-scale magnetic fields \citep[e.g.,][]{2017ApJ...837..121G}.
The ratio between the FIR or the 24~$\mu$m and radio emission also shows a dependence on stellar mass, as also shown by \cite{2021A&A...647A.123D}, who found $q_{\rm IR} \propto (0.15 \pm 0.01) ~ {\rm log} ~ M_\star$ comparable to the SFRS $q_{\rm IR} \propto (0.09 \pm 0.02) ~ {\rm log} ~ M_\star$.
However this correlation is weaker in comparison to the \Ha\ emission, where $q_{\rm H\alpha} \propto 0.22 \pm 0.04 ~ {\rm log} ~ M_\star$.
This difference can be attributed to the inability of IR emission to follow SFR in low-mass dust-deficient galaxies.

\subsection{Contribution of AGNs in SFR measurements and calibrations}
\label{sec:Disc+AGN}

Because the SFRS sample was blindly selected regarding the presence of AGN, it gives a representative picture of the presence of AGN in the overall population of star-forming galaxies in the local Universe.
By covering the full range of 60~$\mu$m luminosity present in the local Universe, it includes normal galaxies, low-luminosity AGN, and even a few well known quasars (e.g., galaxies 3C~273, and OJ~287).
The SFRS galaxies were classified based on BPT diagrams of their nuclear emission-line ratios \citep[][Section \ref{sec:Sample}]{2018MNRAS.475.1485M}.
This allows us to investigate the possible biases introduced in the different SFR calibrations by the contribution of the AGNs to the relevant SFR tracers.
The SFRS consists of 262 star-forming, 39 Seyfert, 26 Composite, 32 LINER galaxies, and 10 galaxies with unreliable or missing classifications. 
For the present \Ha\ sample, there are 224 star-forming, 28 Seyfert, 22 composite, and 26 LINER galaxies.
The inferred SFRs of the Seyfert, composite, and LINER galaxies will mistakenly include an AGN component.

\begin{figure}
    \centering
    \includegraphics[width=\columnwidth]{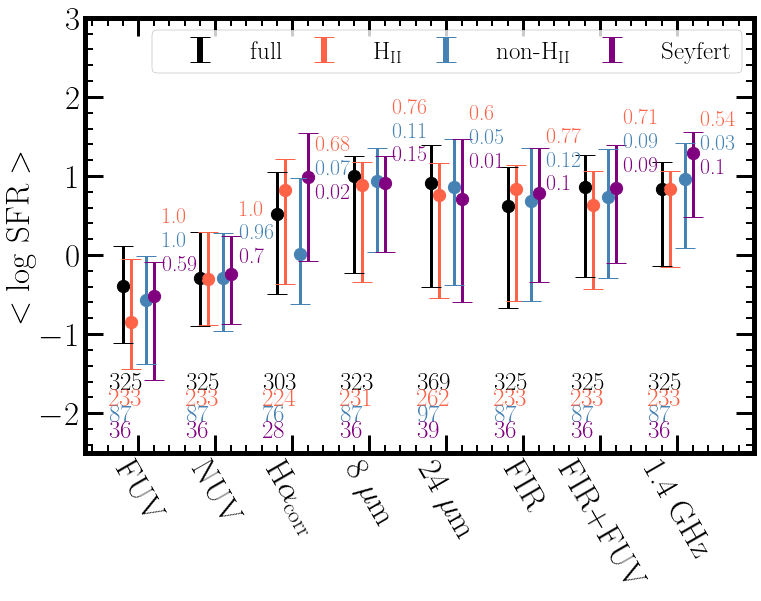}
    \includegraphics[width=\columnwidth]{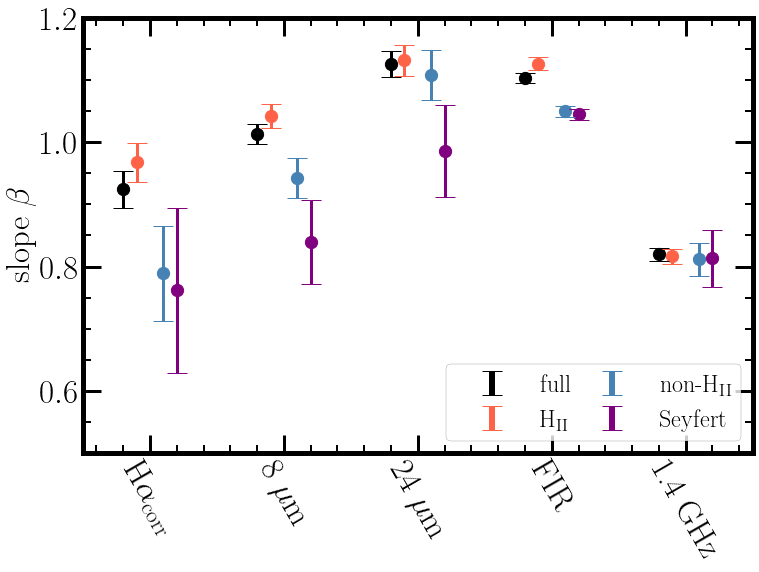}
    \caption{Top: Modes and 68\% confidence intervals of the distribution of inferred SFRs of the complete SFRS (black), \ion{H}{II} (red), non-\ion{H}{II} (blue), and Seyfert (purple) galaxies for various SFR indicators.
    In the same column with the points are the numbers of galaxies in each galaxy class, which may differ based on each indicator's photometric coverage.
    Next to the points are $p$-values from the KS tests comparing the SFR distributions of each class with the full SFRS sample. 
    In each case the colours of the numbers match the colour scheme of the galaxy classes.
    Bottom: The slopes $\beta$ of linear-regression fits of the form $\rm log ~ SFR_{x} = \alpha + \beta~log SFR_{tot}$ for different SFR indicators and each galaxy class.}
    \label{fig:AGNs}
\end{figure}

Figure \ref{fig:AGNs} shows the modes and 68\% percentiles of the distributions of SFRs of the SFRS galaxies as inferred from different SFR indicators for four different samples: all SFRS galaxies (full sample), the star-forming galaxies (\ion{H}{II}) considered so far in this paper, the non-\ion{H}{II} (i.e., Seyfert, Composite, and LINER), and Seyfert galaxies.
The relative number of AGN or non star-forming galaxies slightly differed between the SFR indicators we consider due to variations in the photometric coverage for the different bands. 
The relative numbers are given in the bottom of the top panel of Figure \ref{fig:AGNs}.
The SFR distribution modes are not significantly biased by the presence of AGN-hosting galaxies.
A KS test was performed in order to obtain a picture of the differences between the inferred SFR distributions of the considered subsamples and the full sample.
This way we examine how much the distribution of SFRs in a survey blind to activity classification is affected by AGN galaxies.
The corresponding $p$-values are shown in Figure \ref{fig:AGNs}.
The KS tests suggest that only for the 24~$\mu$m and \Ha\ emission can we reject the null hypothesis that the SFR distributions of the Seyfert galaxies and the full sample are drawn from the same parent distribution, and with less confidence for the non-\ion{H}{II} versus the full sample.
The \ion{H}{II} and full sample galaxies show similar distributions for all SFR indicators.
However, because the full sample contains the individual subsets we are considering, the KS tests can only be considered as indicative.

Overall, this comparison shows that for all the SFR indicators except the 1.4~GHz emission, the bias introduced from AGNs in the average SFR of samples blind to galaxy classification is not significant.
Due to the relative rarity of the luminous AGN in the local Universe, those objects do not dominate the luminosity distribution of galaxies, resulting to small bias in the corresponding SFRs.
This has the important implication that galaxy surveys that do not screen AGN are not necessarily biased towards higher SFR both in terms of their overall statistics and the inferred SFR distributions.

However, SFR measures for individual galaxies can be affected by an AGN, which in turn may bias calibrations of SFR indicators.
Figure \ref{fig:AGNs} shows the slopes of the linear-regression fits between various SFR indicators with $\rm SFR_{tot}$.
In all cases except for the 1.4~GHz emission, blindly including galaxies hosting AGNs in the samples tends to flatten the slopes with the largest discrepancy in the case of Seyfert galaxies.
However, the \ion{H}{II} and the full samples (which would be used in a blind survey) give consistent slopes within the uncertainties.
This is attributed to the fact that SFRS is a FIR selected sample, which is biased towards star-forming galaxies, omitting the bulk of luminous AGN \citep[e.g.,][]{2014ARA&A..52..373L}.

The AGN sample tends to show flatter slopes in the $\rm SFR_{H\alpha}$--$\rm SFR_{tot}$ relation with respect to the \ion{H}{II} and full samples.
This is driven by an \Ha\ excess in the AGN sample in low luminosity galaxies.
The flatter relations of the 24~$\mu$m and FIR-based SFRs with respect to the $\rm SFR_{tot}$ could be due to the generally bluer IR SEDs of AGN in comparison to star-forming galaxies \citep[e.g.,][]{1992MNRAS.254...30G}.
In the case of PAH~8~$\mu$m SFR, the flatter slope with respect to the $\rm SFR_{tot}$ is driven by the preferential destruction of PAH molecules in the intense UV field of luminous AGN.
The calibrations between $\rm SFR_{1.4~GHz}$ and $\rm SFR_{tot}$ is unaffected by AGNs.
This could be because the majority of AGN in our sample are low-luminosity AGN as indicated by the comparison of their luminosity with the nuclear \Ha\ luminosities of star-forming galaxies \citep{2018MNRAS.475.1485M}.
In addition, the radio luminosity distributions of AGN and \ion{H}{II} galaxies \citep[][]{2019MNRAS.482..560M} show that AGNs contribute in the higher luminosity systems, where the contrast with star-forming activity is smaller.
Instead, in the low star-formation regime, the contribution of AGNs is minimal resulting in similar behaviors between the  $\rm SFR_{1.4~GHz}$ and $\rm SFR_{tot}$. 
The slopes of the one-to-one comparisons are more sensitive to the AGN contribution in low-luminosity galaxies, while the modes of the distributions are sensitive to the bulk of the population (most at intermediate luminosities).

\subsection{Metallicity and extinction}
\label{sec:Discussion_Metal}

Given that the dust is composed of metals, one would expect a relation between metallicity and extinction. 
Such a positive correlation has been reported in previous studies  \citep[e.g.,][]{2009ApJ...706..553B,2019ApJ...871..128T,2020ApJ...899..117S,2020ApJ...903L..28S}, although they show significant scatter.

\cite{2018MNRAS.475.1485M} measured the nuclear metallicities for all star-forming galaxies in the SFRS sample. 
For the galaxies with available long-slit spectra, \cite{2018MNRAS.475.1485M} also provided metallicities from large-aperture extractions encompassing the major axis of the galaxy.
We adopted these metallicities calculated through the O3N2 calibration of \cite{2004MNRAS.348L..59P}: 
\begin{eqnarray}
    \rm [12 + log(O/H)] =  8.73 - 0.32 \times O3N2 \quad ,
\end{eqnarray}
where
\begin{eqnarray}
   \rm O3N2 = log \, \frac{\textit{f}_{[O\,_{III}]_{\lambda \, 5007}}/\textit{f}_{H\beta}} {\textit{f}_{[\ion{N}{II}]_{\lambda \, 6583}}/\textit{f}_{H\alpha}} \quad ,
\end{eqnarray}
and \textit{f} corresponds to each emission-line flux.

\begin{figure*}
    \centering
    \includegraphics[width=\linewidth]{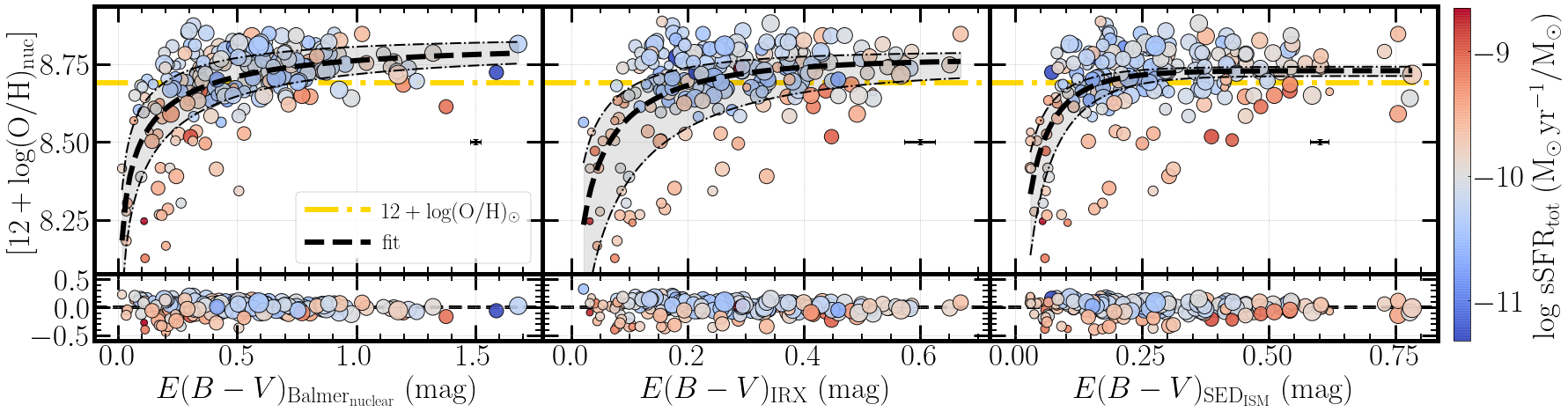}
    \caption{Metallicity of the SFRS star-forming galaxies as a function of extinction. 
    Plots from left to right use Balmer decrement, IRX, and SED-based extinction, respectively.
    The points are color-coded depending on their sSFR, while their size reflects the logarithm of the stellar mass of the galaxies (larger points indicate larger mass; See Figure \ref{fig:PAH_tot} for reference).
    This comparison involves 250, 233, and 221 star-forming SFRS galaxies for the comparison with extinction derived through the Balmer decrement, IRX, and SED fits respectively, depending on spectroscopic and photometric availability, and the quality of the SED fits.
    The yellow dashed-dotted line shows the solar metallicity.
    The black dashed curve shows the best-fit relation (Table \ref{tab:metal_ext}) and the shaded areas indicate the fit uncertainties.
    The black error bar at the center-right of each panel indicates the median uncertainties.
    Subplots show the fit residuals.
    }
    \label{fig:Metal}
\end{figure*}

The metallicities of the SFRS galaxies range from sub-solar to slightly super-solar values ($\rm 8.1 \lesssim [12 + log(O/H)_{nuc}] \lesssim 9$) having  continuous coverage in between (Figure \ref{fig:Metal}). 
The selection of the SFRS galaxies was blind to metallicity, and the SFRS does not represent the full distribution of nearby galaxies' metallicities.
However, it does give a good representation of the metallicity distribution of common star-forming galaxies in the local Universe.

\begin{figure}
    \centering
    \includegraphics[width=0.7\columnwidth]{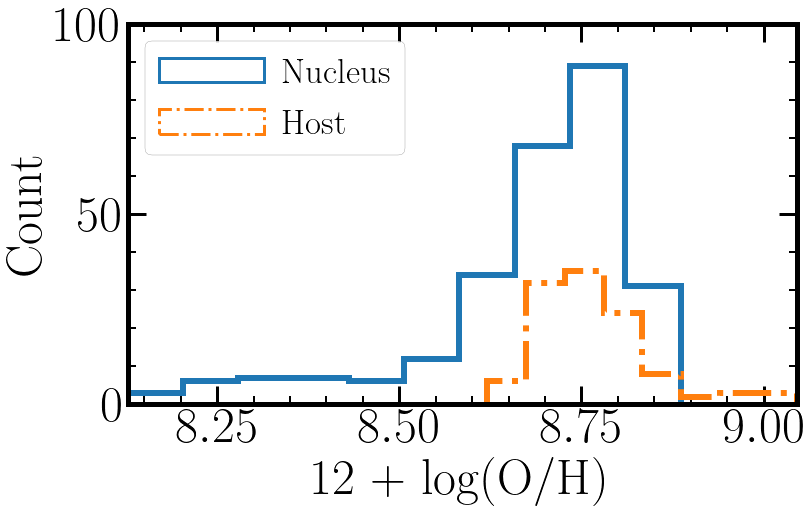}
    \caption{The distribution of metallicities [12 + log(O/H)] for the host and nucleus of SFRS galaxies \protect\citep{2018MNRAS.475.1485M} with orange and blue continuous dashed-dotted lines respectively.}
    \label{fig:Metal_dist}
\end{figure}

Figure \ref{fig:Metal} presents the SFRS galaxies metallicities as a function of the extinction, calculated using each method we considered.
We only considered metallicities derived from the nuclear region because the sample with metallicities tracing larger regions is much smaller and biased to higher-metallicity galaxies (Figure \ref{fig:Metal_dist}).
Nonetheless, these are indicative of the average metallicities and can be used to derive general correlations.
For consistency, in this comparison we use the nuclear Balmer-based extinction, in order to compare with the metallicities derived from the same apertures.

The nuclear-region metallicities of the SFRS galaxies show a non-linear behavior with respect to extinction.
Galaxies with low extinction strongly correlate with low metallicities.
Such galaxies are predominantly low-mass dwarf galaxies, while galaxies with extinction $E(B{-}V)_{\rm Balmer} \gtrsim 0.5 $ have converged to the peak average metallicity of our epoch $\rm [12 + log(O/H)_{nuc}] \simeq 8.75$.
This behavior is similar for the considered extinction indicators, although the range of $E(B{-}V)$ is different.
The comparison with the Balmer-decrement $E(B{-}V)$ shows a tighter relation because metallicity is measured from the same nuclear regions, while the IRX and SED extinctions refer to integrated-galaxy emission.

Unsurprisingly, low-mass galaxies tend to have lower metallicity and extinction. 
This is the result of the well-studied mass/metallicity relation \citep[e.g.,][]{1979A&A....80..155L,2004ApJ...613..898T,2015ARA&A..53...51S,2020MNRAS.491..944C,10.1093/mnras/stab1213}, which reflects the fact that low-mass galaxies have yet to build their stellar component and thus the metal content of their ISM.
Therefore, the power component in Equation \ref{eq:metal_ext} describes the correlation between extinction and metallicity for the young galaxies that are still in the process of building their stellar mass as well as their dust component.
However, intermediate and larger galaxies, have already reached the average peak metallicity of our epoch, and as a result they have increased dust mass, resulting in higher extinction.

The extinction--metallicity correlation can be described by a functional form similar to that presented for the mass--metallicity relation \citep{2020MNRAS.491..944C}:
\begin{eqnarray}
    \label{eq:metal_ext}
    {\rm [12 + log(O/H)]} = C + {\rm log} (1 - 10^{{-\frac{E}{E_0}}^\gamma}) \quad ,
\end{eqnarray}
where $E$ is the extinction, 
$C$ is the asymptotic value of the [12 + log(O/H)] metallicity after converging to the linear part of the correlation, and
$E_0$ is the extinction at the turn-over of the relation. 
This model was fitted with a MCMC using the \texttt{Python emcee} package \citep[][]{emcee}.
The best-fit results for the different extinction indicators are given in Table \ref{tab:metal_ext}.
Despite the significant scatter in the data, Equation \ref{eq:metal_ext} describes quite well the correlation as seen from the residuals plot. 
However, there is a group of points with $E(B{-}V)_{\rm Balmer}< 0.5$ that tends to deviate from this relation. 
This group of points (best seen in the residuals plot at the bottom panels of Figure \ref{fig:Metal}) share the common characteristic of being dwarf highly star-forming galaxies ($\rm sSFR \gtrsim 10^{-9.5} \, M_{\odot} yr^{-1}/M_{\odot}$).
In dwarf galaxies experiencing intense star-formation, the bulk of the \Ha\ emission is expected to originate from individual star-formation sites rather than their main body. 
Therefore, the measured extinction also reflects attenuation by dust in the birth clouds rather than the general ISM. 
Although these galaxies have lower overall dust content, the larger optical depth towards these sites of star formation results in higher measured extinction with respect to other galaxies of the same metallicity. 
In addition, these galaxies are relatively small in size, resulting in larger coverage of the star-forming regions by the general ISM (Section \ref{sec:Discussion_Ha_NII}) and therefore, larger extinction.
The latter is also supported by the fact that, with the use of the IRX extinction indicator (which reflects emission from the full body of the galaxy and not only the nuclear region), these galaxies show even larger differences with respect to the overall relation.

\begin{table}
    \centering
    \caption{Parameters for Eq. \ref{eq:metal_ext} describing the correlation between extinction and metallicity.}
    \label{tab:metal_ext}
    \begin{tabular}{cccc}
    Extinction indicator & $C$  & $E_0$ & $\gamma$ \\
    \hline\\[-4pt]
    Balmer, nuclear & $8.81^{+0.02}_{-0.03}$ & $0.98^{+0.34}_{-0.31}$ & $0.52^{+0.06}_{-0.04}$\\[5pt]
    IRX & $8.76^{+0.03}_{-0.02}$ & $0.28^{+0.13}_{-0.07}$ & $0.74^{+0.12}_{-0.11}$\\[5pt]
    SED ISM & $8.73^{+0.02}_{-0.01}$ & $0.16^{+0.04}_{-0.03}$ & $0.90^{+0.16}_{-0.15}$
    \end{tabular}
\end{table}

\section{Summary}
\label{sec:Summary}
Through the use of the SFRS, a representative sample of local Universe star-forming galaxies this paper has:
\begin{enumerate}
    \item provided \Ha\ photometry for 305 SFRS galaxies
    \item provided calibrations of \Ha-based SFRs with $\rm SFR_{tot}$ using extinction corrections based on the Balmer decrement, IRX, and SED fits, as well as corrections for the contribution of the [\ion{N}{II}] emission.
    \item compared the hybrid indicators of the 24~$\mu$m + \Ha, 8~$\mu$m + \Ha, and FIR~+~FUV, finding good agreement.
    \item shown that SFRs derived through SED fits agree with the SFRs based on FIR~+~FUV emission for $\rm SFR \gtrsim 1 ~ M_\odot \, yr^{-1}$ but show larger scatter for lower SFR.
    \item proposed a new calibration for measuring SFRs from the radio 1.4~GHz emission, based on comparison with \Ha\ emission
    \item shown that low mass galaxies show a deficit in their radio emission with respect to \Ha\ emission
    \item shown that AGNs bias the calibration of SFR indicators but have a small effect when measuring the star-forming activity of large samples of galaxies
    \item provided a function that describes the correlation between the nuclear-region metallicity with the IRX, Balmer decrement, and SED-based extinction for a wide range of metallicity and extinction.
\end{enumerate}

\section*{Acknowledgements}

K. K. would like to thank Ioanna Leonidaki, Maria Kopsacheili, Jeff Andrews, John Kypriotakis, Elias Kyritis, Grigoris Maravelias, Tasos Kougentakis, Anna Steiakaki, and Vangelis Pantoulas for assistance with the \Ha\ observations. 
K. K., and A. Z. acknowledge funding from the European Research Council under the European Union's Seventh Framework Programme (FP/2007-2013)/ERC Grant Agreement n. 617001 (A-BINGOS). 
This project has received funding from the European Union's Horizon 2020 research and innovation programme under the Marie Sklodowska-Curie RISE action, grant agreement No 691164 (ASTROSTAT). 
This research has made use of: (a) software provided by the CXC in the application packages DS9;
(b) data products from the Wide-field Infrared Survey Explorer (WISE), which is a joint project of the University of California, Los Angeles, and JPL, California Institute of Technology, funded by NASA; 
(c) observations made with the Spitzer Space Telescope, which was operated by JPL, California Institute of Technology under a contract with NASA; 
(d) the NASA/IPAC Extragalactic Database (NED), which is operated by the Jet Propulsion Laboratory (JPL), California Institute of Technology, under contract with NASA; 
(e) the NASA/IPAC Infrared Science Archive (IRSA), which is funded by NASA and operated by the California Institute of Technology; 
(f) IRAF which was distributed by the National Optical Astronomy Observatory, which was managed by the Association of Universities for Research in Astronomy under a cooperative agreement with the National Science Foundation.

\section*{DATA AVAILABILITY}

The data underlying this article are available in the article and in its online supplementary material.




\bibliographystyle{mnras}
\bibliography{main}



\section*{Appendix}
\label{sec:Appendix}

General properties of the SFRS galaxies along with the \Ha\ photometry, extinction, metallicities, [\ion{N}{II}]/H$\alpha$ emission ratio, and star-formation rates based on \Ha, PAHs 8~$\mu$m, and 24~$\mu$m emission.
The full table can be found in the electronic form of the article.

\begin{landscape}
    \begin{tablenotes}
    \item The table columns are:\\ 
    \item (1): SFRS ID.
    \item (2): Common galaxy name. 
    \item (3): Right ascension (J2000).
    \item (4): Declination (J2000).
    \item (5): BPT classification presented by \cite{2018MNRAS.475.1485M}. \ion{H}{II}, Sy, TO, and LNR correspond to star-forming, Seyfert, transition object, and LINER galaxies respectively. 
    \item (6): Distance (Mpc) from \cite{2011PASP..123.1011A}.
    \item (7): ${\rm log}~ f_{\rm H\alpha + [\ion{N}{II}]} ~ ({\rm erg~ s^{-1} ~ cm^{-2}})$.
    \item (8): ${\rm log}~ f_{\rm H\alpha + [\ion{N}{II}]} ~ ({\rm erg~ s^{-1} ~ cm^{-2}})$ 68\% uncertainty.
    \item (9) $f_{\rm WISE-1} ~ {\rm (mJy)}$.
    \item (10) $f_{\rm WISE-2} ~ {\rm (mJy)}$.
    \item (11) $f_{\rm WISE-3} ~ {\rm (mJy)}$.
    \item (12) $f_{\rm WISE-4} ~ {\rm (mJy)}$.
    \item (13): $E(B{-}V)_{\rm Balmer,~nuclear}$ presented by \cite{2018MNRAS.475.1485M}. 
    \item (14): $E(B{-}V)_{\rm Balmer}$ calculated as described in Section \ref{sec:Extinction_indicators}.
    \item (15): $E(B{-}V)_{\rm IRX}$ presented by \cite{2019MNRAS.482..560M}.
    \item (16): $E(B{-}V)_{\rm SED}$ from Maragkoudakis et al. in prep.
    \item (17): log $f_{[\ion{N}{II}]}/f_{\rm H\alpha}$ presented by \cite{2018MNRAS.475.1485M}.
    \item (18): $\rm [12+log(O/H)_{nucleus}]$ presented by \cite{2018MNRAS.475.1485M}.
    \item (19): $\rm [12+log(O/H)_{host}]$ presented by \cite{2018MNRAS.475.1485M}.
    \item (20): $\rm {\rm log}~SFR_{ H\alpha} ~(M_\odot ~ yr^{-1}$). \Ha\ flux is corrected for the [\ion{N}{II}] contribution and extinction based on Balmer decrement.
    \item (21): $\rm {\rm log}~SFR_{\rm tot} ~(M_\odot ~ yr^{-1})$ presented by \cite{2019MNRAS.482..560M}.    
    \item (22): $\rm {\rm log}~ SFR_{SED} ~(M_\odot ~ yr^{-1})$ from Maragkoudakis et al. in prep.
    \item (23): $\rm {\rm log}~ SFR_{24\mu m} ~(M_\odot ~ yr^{-1})$.
    \item (24): $\rm {\rm log}~ SFR_{24\mu m + H\alpha} ~(M_\odot ~ yr^{-1}$). \Ha\ flux is corrected for the [\ion{N}{II}] contribution.
    \item (25): $\rm {\rm log}~ SFR_{PAH ~ 8\mu m + H\alpha} ~(M_\odot ~ yr^{-1})$. \Ha\ flux is corrected for the [\ion{N}{II}] contribution.
    \item (26): ${\rm log}~ M_{\star} ~ ({\rm M_\odot})$ presented by \cite{2017MNRAS.466.1192M}.
    \item
\end{tablenotes}
\label{tab:Full_SFRS}
\setlength{\tabcolsep}{2.5pt} 
\renewcommand{\arraystretch}{1.2} 
\footnotesize
\rowcolors{3}{gray!10}{white!50}
\begin{supertabular}{ *{26}{c} }
(1) & (2) & (3) & (4) & (5) & (6) & (7) & (8) & (9) & (10) & (11) & (12) & (13) & (14) & (15) & (16) & (17)  & (18) & (19) & (20) & (21) & (22) & (23) & (24) & (25) & (26)\\
\hline
1 &  IC\,486 & 08:00:20 & +26:36:49 & Sy & $ 114.4 $ & $ -12.46 $ & $ 0.03 $ & $ 35.9 $ & $ 33.17 $ & $ 72.99 $ & $ 208.56 $ & $ 0.61 $ & $ 0.38 $ & $ 0.25 $ & $ 0.64 $ & $ 0.07 $ &     &     & $ 0.62 $ & $ 0.47 $ & $ 0.41 $ & $ 1.07 $ & $ 1.02 $ & $ 0.59 $ &     \\ 

2 &  IC\,2217 & 08:00:50 & +27:30:02 & $\ion{H}{II}$ & $ 76.1 $ & $ -11.86 $ & $ 0.01 $ & $ 18.84 $ & $ 13.13 $ & $ 100.91 $ & $ 235.14 $ & $ 0.47 $ & $ 0.12 $ & $ 0.21 $ & $ 0.61 $ & $ -0.42 $ & $ 8.77 $ & $ 8.72 $ & $ 0.73 $ & $ 0.61 $ & $ 0.71 $ & $ 0.58 $ & $ 0.82 $ & $ 0.9 $ & $ 10.48$ \\ 
3 &  NGC\,2500 & 08:01:53 & +50:44:14 & LNR & $ 15.0 $ & $ -11.66 $ & $ 0.01 $ & $ 79.51 $ & $ 46.32 $ & $ 128.75 $ & $ 193.41 $ & $ 0.07 $ & $ 0.06 $ & $ 0.07 $ & $ 0.15 $ & $ -0.45 $ &     &     & $ -0.56 $ & $ -0.28 $ & $ -1.02 $ & $ -0.99 $ & $ -0.51 $ & $ -0.35 $ &     \\ 

4 &  NGC\,2512 & 08:03:08 & +23:23:31 & $\ion{H}{II}$ & $ 69.3 $ & $ -12.19 $ & $ 0.14 $ &     &     &     &     & $ 0.43 $ & $ 0.37 $ &     &     & $ -0.31 $ & $ 8.85 $ & $ 8.74 $ & $ 0.62 $ &     &     & $ 0.81 $ & $ 0.8 $ &     & $ 10.9$ \\ 
5 &  MCG\,6-18-009 & 08:03:29 & +33:27:44 & $\ion{H}{II}$ & $ 164.3 $ &     &     & $ 19.98 $ & $ 12.89 $ & $ 56.69 $ & $ 141.8 $ & $ 0.52 $ &     & $ 0.28 $ & $ 0.71 $ & $ -0.29 $ & $ 8.8 $ &     &     & $ 1.03 $ & $ 1.07 $ & $ 1.17 $ &     &     & $ 11.24$ \\ 
6 &  MK\,1212 & 08:07:06 & +27:07:34 & $\ion{H}{II}$ & $ 173.3 $ &     &     &     &     &     &     & $ 0.66 $ &     &     &     & $ -0.28 $ & $ 8.74 $ &     &     &     &     & $ 1.31 $ &     &     & $ 10.95$ \\ 
7 &  IRAS\,08072+1847 & 08:10:07 & +18:38:18 & $\ion{H}{II}$ & $ 70.8 $ & $ -13.25 $ & $ 0.02 $ &     &     &     &     & $ 0.9 $ & $ 0.59 $ &     &     & $ -0.2 $ & $ 8.74 $ &     & $ -0.17 $ &     &     & $ 0.88 $ & $ 0.78 $ &     & $ 10.01$ \\ 
8 &  NGC\,2532 & 08:10:15 & +33:57:24 & $\ion{H}{II}$ & $ 77.6 $ & $ -11.69 $ & $ 0.01 $ & $ 60.59 $ & $ 39.97 $ & $ 274.22 $ & $ 549.14 $ & $ 0.85 $ & $ 0.34 $ & $ 0.19 $ & $ 0.42 $ & $ -0.42 $ & $ 8.76 $ & $ 8.72 $ & $ 1.21 $ & $ 0.93 $ & $ 0.96 $ & $ 0.89 $ & $ 1.07 $ & $ 1.19 $ & $ 11.11$ \\ 
9 &  UGC\,4261 & 08:10:56 & +36:49:41 & $\ion{H}{II}$ & $ 93.2 $ & $ -12.34 $ & $ 0.07 $ & $ 8.05 $ & $ 5.31 $ & $ 30.56 $ & $ 112.35 $ & $ 0.33 $ & $ 0.05 $ & $ 0.13 $ & $ 0.43 $ & $ -0.6 $ & $ 8.53 $ &     & $ 0.37 $ & $ 0.54 $ & $ 0.46 $ & $ 0.79 $ & $ 0.83 $ & $ 0.6 $ & $ 10.14$ \\ 
10 &  NGC\,2535 & 08:11:13 & +25:12:24 & $\ion{H}{II}$ & $ 61.6 $ & $ -11.83 $ & $ 0.09 $ & $ 39.01 $ & $ 22.81 $ & $ 126.37 $ & $ 265.92 $ & $ 0.32 $ & $ 0.15 $ & $ 0.17 $ & $ 0.45 $ & $ -0.43 $ & $ 8.87 $ & $ 8.66 $ & $ 0.63 $ & $ 0.52 $ & $ 0.59 $ & $ 0.34 $ & $ 0.63 $ & $ 0.75 $ & $ 10.68$ \\
\end{supertabular}
\end{landscape}

\bsp	
\label{lastpage}
\end{document}